%% file: arxiv_v1.tex
\documentclass[a4paper,11pt]{article}

\usepackage{graphicx,array}
\usepackage{url}
\usepackage{color}
\usepackage{latexsym}
\usepackage{amsthm}
\usepackage{braket}
\usepackage{amsmath}
\usepackage{amsfonts}
\usepackage{amssymb}
\usepackage{mathtools}
\usepackage{bbold}
\usepackage{amsmath,amssymb}
\usepackage{slashed}
\usepackage{mathrsfs}
\usepackage{enumerate}
\usepackage{tikz}
\usepackage{mdframed}
\usepackage{setspace}  
\usepackage{esvect}
\usepackage{esint}

\usepackage{subcaption}
\usepackage[normalem]{ulem}
\usepackage{setspace}
\usepackage{multirow,makecell}

\usepackage[utf8x]{inputenc}%
\usepackage{tcolorbox}%

\usepackage{jheppub} 
\usepackage{orcidlink} 
\usepackage{etoolbox}
\makeatletter
\renewcommand{\@fpheader}{}
\patchcmd{\maketitle}
  {\noindent{\textsc{Abstract:}}\ \@abstract}
  {\noindent\@abstract}
  {}{\PackageError{draft_v5}{Could not remove abstract label}{Check the JHEP title format.}}
\makeatother

\usepackage[all]{hypcap} 

\newcommand{\nc}{\newcommand}

\nc{\beq}{\begin{equation}}  
\nc{\eeq}{\end{equation}}  
\nc{\beqa}{\begin{eqnarray}}  
\nc{\eeqa}{\end{eqnarray}}  
\nc{\bit}{\begin{itemize}}  
\nc{\eit}{\end{itemize}}

\usepackage{floatrow}
\newfloatcommand{capbtabbox}{table}[][\FBwidth]

\usepackage{blindtext}

\newcommand{\be}{\begin{equation}}
\newcommand{\ee}{\end{equation}} 
\newcommand{\bea}{\begin{eqnarray}}
\newcommand{\eea}{\end{eqnarray}}
\newcommand{\bc}{\begin{center}}
\newcommand{\ec}{\end{center}}

\def\cale{{\mathcal{E}}}

\newcommand{\ab}[1]{\langle #1 \rangle}

\newcommand{\la}[1]{\langle #1 |}

\newcommand{\ra}[1]{| #1 \rangle }

\newcommand{\En}[1]{\mathcal{E}(#1)}

\def\b{\beta}

\def\cale{{\mathcal{E}}}

\begin{document}
\hypersetup{pageanchor=false} 

\preprint{CERN-TH-2026-302}

\title{All the Positive Energy Correlators}

\author[a]{Lorenzo Ricci~\orcidlink{0000-0001-8704-3545},}
\author[a,b]{Marc Riembau~\orcidlink{0000-0002-9842-2425}}
\author[c]{and Minho Son~\orcidlink{0000-0001-9913-7784}}

\affiliation[a]{Theoretical Physics Department, CERN,\\ 1211 Geneva 23, Switzerland}
\affiliation[b]{Theoretical Particle Physics Laboratory, Institute of Physics, EPFL,\\ Lausanne, Switzerland}
\affiliation[c]{Department of Physics, Korea Advanced Institute of Science and Technology,\\ 291 Daehak-ro, Yuseong-gu, Daejeon 34141, Republic of Korea}

\abstract{
We develop a framework to systematically obtain the full set of positivity bounds on multi-point energy correlators for sources of arbitrary spin. The constraints are organized in a family of positive operators $\mathscr{E}^{(N,k)}$, with $k=0,\ldots,\lfloor N/2\rfloor$. 
Our approach provides outer bounds, which exclude correlators that violate the positivity of $\mathscr{E}^{(N,k)}$, and inner bounds, obtained by constructing ensembles of consistent collider events via linear programming. 
Both methods converge to each other as numerical precision increases. 
This approach, in particular, characterizes the allowed higher-point correlators once lower-point correlators are determined. 
We illustrate this through several examples, including bounds on multipoles of the three-point energy correlator in QCD derived from measurements of one- and two-point correlators at LEP.
}

\notoc 
\maketitle

\thispagestyle{empty}  
\newpage  
  
\setcounter{page}{1}
\hypersetup{pageanchor=true}

\begingroup
\hypersetup{linkcolor=black,linktocpage}
\tableofcontents
\endgroup
\newpage


\section{Introduction}\label{Sec:intro}

Collider experiments can be characterized by the correlations of the asymptotic energy flux generated. These observables, dubbed Energy Correlators (ECs), are defined via the expectation value of idealized energy detectors $\En{\vec{n} }$ in the background generated by some source $\mathcal O$, 
\begin{align}
\braket{\En{\vec{n}_1}\dots \En{\vec{n}_N}}\,\equiv\,\frac{1}{\mathcal N}\int d^4x e^{ip\cdot x} \langle 0| \mathcal O^\dagger(x) \En{\vec{n}_1}\dots \En{\vec{n}_N} \mathcal O(0)|0\rangle\,,
\end{align}
normalized by the norm or total rate $\mathcal N = \int d^4x e^{ip\cdot x} \langle 0| \mathcal O^\dagger(x)  \mathcal O(0)|0\rangle$. 
When asymptotic multiparticle states $\ra{\alpha}$ exist, and there is a well defined S-matrix, the detectors $\En{\vec{n}}$ act on those simply by measuring the energy $E_i$ of the particles moving in the direction $\vec n$, $\En{\vec{n}}\ra{\alpha} = \sum_i E_i \delta^{(2)}(\Omega_n-\Omega_i)|\alpha\rangle$.\footnote{Here and throughout the rest of the paper, we neglect finite mass effects of final state particles unless otherwise specified.} Furthermore, even when multiparticle states are ill-defined, as in Conformal Field Theories (CFT), 
the idealized calorimeters $\En{\vec{n}}$ allow for a representation in terms of the stress energy tensor integrated along a null-ray at infinity \cite{Sveshnikov:1995vi,Belitsky:2001ij,Lee:2006fn,Hofman:2008ar,Kravchuk:2018htv}. Energy correlators were introduced to characterize hadronic events~\cite{Basham:1977iq} and have recently attracted considerable attention both in collider applications and in conformal field theories,  see \cite{Moult:2025nhu} for a review of recent developments.

An intuitive, yet subtle, property of energy detectors $\En{\vec{n}}$ is that they are positive-semidefinite and mutually commuting operators \cite{PhysRevD.43.2542,PhysRevD.44.403,Faulkner:2016mzt,Hartman:2016lgu,Kologlu:2019bco}. Besides having nontrivial implications for CFT data \cite{Hofman:2016awc,Belin:2019mnx,Komargodski:2016gci,Cordova:2017dhq,Costa:2017twz,Meltzer:2018tnm} and renormalization-group flows \cite{Hartman:2023qdn,Hartman:2023ccw}, these properties imply the positivity of ECs for arbitrary detector configurations, namely
\begin{align}\label{Eq:LocalPositivity}
    \braket{\Psi|\En{\vec{n}_1}\ldots \En{\vec{n}_N}|\Psi} \ge 0\,,
\end{align}
for any positive-energy momentum eigenstate $\ket{\Psi}$ and any configuration of detector directions $\vec n_i$.\footnote{More precisely, strong commutativity together with positivity of detector is sufficient to have \eqref{Eq:LocalPositivity} \cite{Belin:2026wkc}.} In the context of CFTs, constraints of this type are often referred to as \textit{conformal collider bounds} \cite{Hofman:2008ar}. 
As shown in \cite{Belin:2026wkc}, these positivity properties also lead to a representation of ECs as averages over an ensemble of events. 

While this is a trivial observation in theories with an S-matrix, where the ensemble of events or states is the fully differential cross section, it is a non-trivial observation for theories that do not admit an S-matrix description. 
A consequence of this representation is that correlators of energy detectors act as a positive kernel in unitary theories. Therefore, one can construct states $\ket{\Psi'} = \int d\vec{n}_1 \ldots d \vec{n}_M f(\vec{n}_1,\ldots, \vec{n}_M)\En{\vec{n}_1} \ldots  \En{\vec{n}_M} \ket{\Psi}$, which have positive norm and positive energy flux in any direction,
\begin{align}\label{Eq:GlobalPositivity}
    \braket{\Psi'|\En{\vec{n}}| \Psi'} \ge 0&& \forall\,\,\, \vec{n}_i, \,\,\, \ket{\Psi'}\,.
\end{align}
These two properties, referred to as \textit{local} and \textit{global} positivity, respectively, in \cite{Belin:2026wkc}, imply a set of consistency conditions that any EC satisfies. It is worth emphasizing that \eqref{Eq:GlobalPositivity} follows from the operator positivity underlying \eqref{Eq:LocalPositivity}, but it provides additional constraints when local positivity is imposed only up to finite $N$. These constraints are, in general, recovered from local positivity only when the latter is imposed consistently for arbitrarily large $N$.

General constraints implied by \eqref{Eq:LocalPositivity} and \eqref{Eq:GlobalPositivity} have been worked out in several previous works. For spinless sources, an early example is the positivity of the multipoles in \cite{PhysRevLett.41.1581}. More recently, this positivity has been used to constrain CFT data \cite{Mecaj:2025ecl,Mecaj:2026kji,Dempsey:2025yiv}. Ref.~\cite{Belin:2026wkc} showed that local positivity and commutativity imply the existence of a representation in terms of an ensemble of events, allowing the exact allowed region for energy correlators to be determined.
For spinning sources, positivity of the one-point correlator leads to the conformal collider bounds \cite{Hofman:2008ar,Cordova:2017zej}, was applies to QCD data in \cite{Riembau:2025wjc} and extended to higher points in \cite{Riembau:2025isw}. Building on these works, the aim of this work is to provide a unified framework that organizes such constraints for any number of detectors and for sources of any spin, connecting correlators with different numbers of detector.

In this work, we develop a simple and systematic framework for deriving and classifying the complete set of positivity bounds on arbitrary $N$-point ECs, for both spinless and spinning sources. Our approach not only extends the analysis of \cite{Belin:2026wkc} to spinning sources, but also systematically incorporates information from lower-point ECs to derive consistent bounds on the shape of higher-point ECs.

We present two complementary approaches to bound ECs. From one side, we present a systematic way to \textit{rule out} ECs incompatible with \eqref{Eq:LocalPositivity} and \eqref{Eq:GlobalPositivity}. We do so by writing these conditions as a set of operators $\mathscr{E}^{(N,k)}$, whose positive semidefiniteness implies all constraints obtained from $N$ detectors, with $k=0,\dots,\lfloor N/2\rfloor$. 
This is an infinite set of necessary conditions. 
The operators with $k\neq 0$ use global positivity, which, as mentioned, is not a fundamental condition but rather a derived property. Strictly speaking, therefore, the full constraints are encapsulated in $\mathscr{E}^{(N,0)}$ for $N\to\infty$. 
However, in practice, when truncating at finite $N$, the  operators with $k\neq 0$ lead to independent constraints, thereby accelerating convergence. Moreover, the operators at finite $N$ and $k$ give an \textit{outer} bound rigorously excluding inconsistent ECs. On the other hand, we \textit{rule in} ECs via a simple linear program that extremizes an observable subject to a given set of constraints, thereby obtaining an explicit representation of extremal ECs. 
The linear program truncates in the space of energy configurations, and therefore it gives robust \textit{inner} bounds. 
By combining both approaches and studying their convergence, we thus provide a method for obtaining rigorous two-sided bounds on ECs.

The constraints on given $N$-point ECs depend on the values of lower-point correlators and the spin structure of the source. This observation dates back to \cite{Zhiboedov:2013opa}, where it was shown that, in CFTs with extremal values of the Weyl-anomaly coefficients ratio $a/c$, the one-point EC saturates the boundary of its allowed region, forcing higher-point ECs to reduce to their free-theory counterparts. The underlying physical intuition is that when the energy flux is restricted to certain directions on the celestial sphere and vanishes identically in others, the dynamics must exhibit free-theory-like behavior. This intuition can be made precise through a simple argument for a spin-one source ``$\ket{J}$'' in 4D. The one-point EC is given by
\begin{align}\label{eq:onepointintro}
\braket{J^i|\En{\vec{n}}|J^j} \,=\, \frac{E}{4\pi} \left( \delta^{ij} + a_\cale (3 n^i n^j - \delta^{ij}) \right)\,,
\end{align}
where, for simplicity, we are assuming the source at rest and with energy $E$. Positive semidefiniteness of this tensor requires $-\frac12 \leq a_\cale \leq 1$~\cite{Hofman:2008ar}. At the endpoints of this interval, the tensor structure reduces to $n^i n^j$ for $a_\cale=1$ and $\delta^{ij}-n^i n^j$ for $a_\cale=-\frac12$. Their kernels force the underlying events to be collinear or planar, respectively; the event orientations are averaged over. 
Once specified, the one-point EC also constrains the possible two-point fluxes. Take the $a_\cale=1$ case. The one-point tensor has rank one at each direction, so positivity requires the two-point tensor to be proportional to\footnote{In particular, since $\En{\vec{n}}$ is a positive semidefinite operator, then $\braket{\Psi| \En{\vec{n}}|\Psi} =0$ implies $\En{\vec{n}} \ket{\Psi} = 0$.} $\braket{J^i |\En{\vec{n}_1 } \En{\vec{n}_2}| J^j} = F_{1}(\chi) n_1^i n_2^j$ with $F_{1}(\chi)$ a function of the detector separation $\chi = \vec{n}_1 \cdot \vec{n}_2$. Yet, commutativity requires $F_{1}(\chi) (n_1^i n_2^j-n_2^i n_1^j) = 0, \,\,\, \forall \vec{n}_1 \,, \vec{n}_2$ giving $F_1(\chi) =  c (\delta(\vec{n}_1-\vec{n}_2)-\delta(\vec{n}_1+\vec{n}_2))$ with $c$ set by energy and momentum conservation. Thus, at $a_\cale=1$, every underlying event has the energy flow of two back-to-back particles. 
The other extreme is less trivial. For noncollinear directions, the two-point correlator takes the form $\braket{J^i |\En{\vec{n}_1 } \En{\vec{n}_2}| J^j} \propto F_{-\frac{1}{2}}(\chi) (\vec{n}_1 \times \vec{n}_2)^i (\vec{n}_1 \times \vec{n}_2)^j$ with possible contact terms on collinear configurations. The underlying events are planar, with their normals correlated with the source weights.\footnote{This is indeed the case since both a 2-particle free fermion state and the Wess-Zumino-Witten current saturate the bound \cite{Riembau:2025isw}.} 
This argument illustrates that positivity and consistency connect different ECs. Indeed, at $a_\cale= 1$, a single lower point spinning observable fully determines higher point ECs.
The presented formalism fully generalizes and systematizes these arguments.

\medskip

The paper is structured as follows. In Section~\ref{Sec:OuterMoments}, we construct the complete set of positive operators $\mathscr{E}^{(N,k)}$ in both 3D and 4D, which yield the \textit{outer} bounds. 
In Section~\ref{Sec:InnerNumerics}, we present the linear program that constructs extremal solutions to a given observable for any set of constraints, providing our \textit{inner} bounds. We illustrate both methods in Section~\ref{Sec:results} and demonstrate their convergence through several case studies.

\section{Excluding Energy Correlators: all the constraints}\label{Sec:OuterMoments}

Given a state $\ket{\Psi}$, local positivity \eqref{Eq:LocalPositivity}, global positivity \eqref{Eq:GlobalPositivity} and energy and momentum conservation impose an infinite tower of constraints on energy correlators. In what follows, we derive the complete set of such constraints.

\subsection{Energy detectors in 3D}\label{sec:3d}
In 3D, the celestial sphere is a circle $S^1$, and $\vec{n}$ is parametrized by a single angle $\theta\in [0,2 \pi)$. To find the complete set of constraints it is convenient to perform a mode decomposition and define the operators
\begin{align}\label{Eq:modes3d}
\hat{e}_n\,=\,\int d\theta \, e^{-in\theta}\cale(\theta)\,, \qquad n\in\mathbb{Z}\,.
\end{align}
Since $\cale(\theta)$ is hermitian, $\hat{e}_n^{\dagger} = \hat{e}_{-n}$, so the independent modes are those with $n \ge 0$. 
Parity acts as $\mathsf P\hat e_n\mathsf P^{-1}=\hat e_{-n}$; it does not identify these operators. Their expectation values are real in a parity-invariant state. 
Measuring energies in units of the total energy $E$, energy conservation implies $\hat{e}_0 = \mathbb{1}$, while momentum conservation, assuming massless particles, gives $\hat{e}_{1} = \hat{e}_{-1} = 0$. 
They mean that energy and momentum are conserved exactly, not in average, so both hold as operator identities on the states we consider, and not only as expectation values.

Let us now turn to positivity. We assume throughout that energy detectors at distinct angles commute, so that all the operators $\hat{e}_n$ commute among themselves. That energy is positive for any angle, i.e. $\cale(\theta)\ge 0$ for every $\theta$, implies that the integral of the energy against any positive function is positive. By writing an arbitrary function as $p(\theta)=\sum_n c_n e^{in\theta}$, we have
\begin{align}
    \int d \theta\,  |p(\theta)|^2\, \cale(\theta) = \sum_{i,j} c_i^*\, c_j\, \hat{e}_{i-j} \,\ge\, 0\,.
\end{align}
As this holds for arbitrary coefficients $c_i$, it is equivalent to the positivity of the matrix
\begin{align}\label{Eq:E1}
    \mathscr{E}^{(1,0)}_{i;j}\equiv \hat{e}_{i-j}\,, \qquad \mathscr{E}^{(1,0)} \succeq 0\,.
\end{align}
This is a positive semidefinite matrix of operators, whose entries are the different modes of the energy flux. 
The statement $\mathscr{E}^{(1,0)}\succeq 0$ means that $\sum_{i,j}\braket{\Psi_i|\mathscr{E}^{(1,0)}_{i;j}|\Psi_j}\ge0$ for any collection of states $\ket{\Psi_i}\in\mathcal{H}$, as follows from $\int d\theta\,\braket{\chi(\theta)|\cale(\theta)|\chi(\theta)}\ge0$ with $\ket{\chi(\theta)}=\sum_n e^{in\theta}\ket{\Psi_n}$. In particular, for $\ket{\Psi_i}=c_i\ket{\Psi}$, the numerical matrix $\braket{\Psi|\mathscr{E}^{(1,0)}_{i;j}|\Psi}$ on $V$ is positive semidefinite for every state $\ket{\Psi}$. Using the Ward identities, the $\mathscr{E}^{(1,0)}$ operator has the explicit form
\begin{align}
    \mathscr{E}^{(1,0)}=\begin{pmatrix}
    1 & 0 & \hat{e}_2^\dagger & \hat{e}_3^\dagger & \dots \\
    0 & 1 & 0 & \hat{e}_2^\dagger & \\
    \hat{e}_2 & 0 & 1 & 0 & \\
    \hat{e}_3 & \hat{e}_2 & 0 & 1 & \\
    \vdots & & & & \ddots\\
\end{pmatrix}
\,\succeq\, 0\,.
\end{align}
This operator contains the full set of constraints obtained from a single insertion of the energy flux. 

For the two point correlator, we have that $\cale(\theta_1)\cale(\theta_2)$ is positive for any pair of angles. 
Then, similarly as in the one-point case, for any function
$f(\theta_1,\theta_2)=\sum_{a,c} c_{ac}\, e^{i(a\theta_1+c\theta_2)}$ we have
\begin{align}
    \int d\theta_1\, d\theta_2\, |f(\theta_1,\theta_2)|^2\, \cale(\theta_1)\cale(\theta_2)
    = \sum_{a,b,c,d} c^*_{ac}\, c_{bd}\; \hat{e}_{a-b}\,\hat{e}_{c-d} \,\ge\, 0\,,
\end{align}
where we used that $\cale(\theta_1)$ and $\cale(\theta_2)$ commute, so that their product is a positive operator. The coefficients $c_{ac}$ are arbitrary, and in particular full rank and need not factorize. The condition is then the positivity of a matrix, with rows labelled by the pair $ac$ and columns by the pair $bd$, and with entries $\hat{e}_{a-b}\,\hat{e}_{c-d}=\mathscr{E}^{(1,0)}_{a;b}\,\mathscr{E}^{(1,0)}_{c;d}$.

This structure will appear repeatedly, with higher point operators inheriting the structure of lower point ones, so we introduce the following notation where given two matrices $A$ and $B$, 
we define the matrix $A\boxtimes B$ by
\begin{align}\label{Eq:boxtimes}
    \left(A\boxtimes B\right)_{ac;bd} \equiv A_{a;b}\, B_{c;d}\,,
\end{align}
where on the right-hand side the entries are multiplied as operators on $\mathcal{H}$. In other words, $\boxtimes$ is a tensor product in the mode indices only: the Hilbert space is not doubled, and both factors act on the same $\mathcal{H}$. In particular $\braket{\Psi|A_{a;b}B_{c;d}|\Psi}\neq\braket{\Psi|A_{a;b}|\Psi}\braket{\Psi|B_{c;d}|\Psi}$ in general, so the matrix of expectation values of $A\boxtimes B$ is not the tensor product of those of $A$ and $B$. Its positivity is a new condition, which we write as
\begin{align}\label{Eq:E2}
    \mathscr{E}^{(2,0)}_{ac;bd} \equiv \left(\mathscr{E}^{(1,0)}\boxtimes\mathscr{E}^{(1,0)}\right)_{ac;bd} = \hat{e}_{a-b}\,\hat{e}_{c-d}\,, \qquad \mathscr{E}^{(2,0)} \,\succeq\, 0\,.
\end{align}
The Ward identities tie some of its entries to
the one-point data: $\hat{e}_0\,\hat{e}_n=\hat{e}_n$ and $\hat{e}_{\pm1}\hat{e}_n=0$, so that \eqref{Eq:E1} is contained in \eqref{Eq:E2} as the block $a=b=0$.

For the two-point correlator we have a second condition, independent of the positivity of $\mathscr{E}^{(2,0)}$, coming from the fact that $\cale(\theta_1)\cale(\theta_2)$ can be used as a positive kernel. 
This is the so-called global positivity condition of \eqref{Eq:GlobalPositivity} or simply unitarity,
\begin{align}\label{eq:globalposN2}
    \int d\theta\, d\theta'\, f^*(\theta) f(\theta')\, \cale(\theta)\, \cale(\theta') \,\ge\, 0\,.
\end{align}
Writing $f(\theta)=\sum_n c_n e^{-in\theta}$, so that $\int d\theta f(\theta)\cale(\theta)=\sum_n c_n \hat{e}_n$, this becomes
\begin{align}\label{Eq:G}
    \sum_{m,n} c_m^*\, c_n\, \hat{e}_m^\dagger \hat{e}_n \ge 0
    \qquad\Longrightarrow\qquad
    \mathscr{E}^{(2,1)}_{m;n}\equiv \hat{e}_{-m}\,\hat{e}_n\,, \qquad \mathscr{E}^{(2,1)} \,\succeq\, 0\,, \qquad m,n\in\mathbb{Z}\,.
\end{align}
So the outer product of the energy modes forms trivially a positive semidefinite matrix. 
Unlike $\mathscr{E}^{(1,0)}$, the entries of $\mathscr{E}^{(2,1)}$ contain two detectors. 
This condition is independent of \eqref{Eq:E2}, as it does not rely on $\cale(\theta)\ge 0$, but only on the hermiticity of $\cale(\theta)$. Again the Ward identities relate it to the one-point data, $\mathscr{E}^{(2,1)}_{0;n}=\hat{e}_n$ and $\mathscr{E}^{(2,1)}_{\pm 1;n}=0$. The two-point correlator leads therefore to two independent positive matrices of operators, $\mathscr{E}^{(2,0)}\succeq0$ and $\mathscr{E}^{(2,1)}\succeq0$.

\medskip

This can be readily extended to products of more detectors. With three detectors there are again two conditions. The first follows from local positivity \eqref{Eq:LocalPositivity} for three detectors, integrated against $|f(\theta_1,\theta_2,\theta_3)|^2$,
\begin{align}\label{Eq:E3}
    \mathscr{E}^{(3,0)}_{abc;def} \equiv \left(\mathscr{E}^{(1,0)}\boxtimes\mathscr{E}^{(1,0)}\boxtimes\mathscr{E}^{(1,0)}\right)_{abc;def} = \hat{e}_{a-d}\,\hat{e}_{b-e}\,\hat{e}_{c-f}\,, \qquad \mathscr{E}^{(3,0)} \succeq 0\,.
\end{align}
The second combines local and global positivity. We treat $\cale(\theta_1)\cale(\theta_2)$ as a positive kernel against any function, so the integral against $g(\theta_1)g^*(\theta_2)$ is positive. This can be extended to any $\theta_3$-dependent function, so
\be\label{eq:globalposN3}
\int d\theta_1d\theta_2d\theta_3 g(\theta_3,\theta_1)g^*(\theta_3,\theta_2) \cale(\theta_1)\cale(\theta_2)\cale(\theta_3) =  \sum_{i,j,m,n} v^*_{im}\, v_{jn}\; \hat{e}_{i-j}\,\hat{e}_{-m}\,\hat{e}_n \,\ge\, 0\,,
\ee
where we expanded the generic function $g$ as $g(\theta,\theta')=\sum_{j,n} v_{jn}\, e^{ij\theta} e^{-in\theta'}$. Since  $v_{jn}$ are full rank, this implies the positive semidefiniteness of the matrix $\mathscr{E}^{(3,1)}$, defined as
\begin{align}\label{Eq:G3}
    \mathscr{E}^{(3,1)}_{im;jn} \equiv \left(\mathscr{E}^{(1,0)}\boxtimes\mathscr{E}^{(2,1)}\right)_{im;jn} = \hat{e}_{i-j}\,\hat{e}_{-m}\,\hat{e}_n\,, \qquad \mathscr{E}^{(3,1)} \succeq 0\,.
\end{align}
Again, this condition is independent of the positivity of $\mathscr{E}^{(3,0)}$. The condition is equivalent to constructing some general $\theta$-dependent combination of operator modes as $A(\theta)=\int d\theta'\, g(\theta,\theta')\,\cale(\theta') = \sum_{j,n} v_{jn}\, e^{ij\theta}\, \hat{e}_n$, and demanding the positivity of $\int d \theta\, A^\dagger(\theta)\, \cale(\theta)\, A(\theta) $.

\medskip

The general pattern is now clear. Every positive matrix is built out of factors of $\mathscr{E}^{(1,0)}$ and $\mathscr{E}^{(2,1)}$. With $N$ detectors and $k$ factors of $\mathscr{E}^{(2,1)}$,
\begin{align}\label{Eq:ENk}
    \mathscr{E}^{(N,k)} \equiv \underbrace{\mathscr{E}^{(1,0)} \boxtimes \dots \boxtimes \mathscr{E}^{(1,0)}}_{N-2k} \boxtimes \underbrace{\mathscr{E}^{(2,1)} \boxtimes \dots \boxtimes \mathscr{E}^{(2,1)}}_{k}\,, \qquad \mathscr{E}^{(N,k)} \succeq 0\,,
\end{align}
which is a positive semidefinite tensor in the sense of being positive when contracted with arbitrary tensors with $N-k$ mode indices, and whose entries are given by
\begin{align}
    \mathscr{E}^{(N,k)}_{a_1\ldots a_{N-2k}\, m_1\ldots m_k\,;\, b_1\ldots b_{N-2k}\, n_1\ldots n_k} = \prod_{r=1}^{N-2k}\hat{e}_{a_r-b_r}\,\prod_{l=1}^{k}\hat{e}_{-m_l}\,\hat{e}_{n_l}\,.
\end{align}
The labels simply add under the product, $\mathscr{E}^{(N,k)}\boxtimes\mathscr{E}^{(N',k')}=\mathscr{E}^{(N+N',k+k')}$. Positivity follows as above from
\begin{align}
    \int d\theta_1\ldots d\theta_{N-2k}\; A^\dagger\, \cale(\theta_1)\cdots\cale(\theta_{N-2k})\, A\,\ge\,0\,,
\end{align}
with $A(\theta_1,\ldots,\theta_{N-2k})=\sum v_{b_1\ldots n_k}\, e^{i(b_1\theta_1+\ldots)}\,\hat{e}_{n_1}\cdots\hat{e}_{n_k}$ and arbitrary coefficients $v$. The $N$-point correlator is then constrained by the $\lfloor N/2\rfloor+1$ conditions
\begin{align}
    \mathscr{E}^{(N,k)}\succeq 0\,, \qquad k=0,\ldots,\lfloor N/2\rfloor\,,
\end{align}
so that the number of independent conditions grows with the number of detectors, as summarized in Table~\ref{tab:ENk}.

\begin{table}[htbp]
    \centering
    \begin{tabular}{c c}
        \hline
        $N$ & Positive matrices $(N,k)$ \\
        \hline
        1 & $(1,0)$ \\
        2 & $(2,0),\ (2,1)$ \\
        3 & $(3,0),\ (3,1)$ \\
        4 & $(4,0),\ (4,1),\ (4,2)$ \\
        5 & $(5,0),\ (5,1),\ (5,2)$ \\
        6 & $(6,0),\ (6,1),\ (6,2),\ (6,3)$ \\
        $\vdots$ & $\vdots$ \\
        \hline
    \end{tabular}
    \caption{Positive matrices $\mathscr{E}^{(N,k)}$ constraining the $N$-point energy correlator. 
    }
    \label{tab:ENk}
\end{table}

Because of the Ward identities these conditions are nested. Setting $a_r=b_r=0$ in one of the $\mathscr{E}^{(1,0)}$ factors gives $\hat{e}_0=\mathbb{1}$, so that $\mathscr{E}^{(N-1,k)}$ is a block of $\mathscr{E}^{(N,k)}$; setting $m_l=n_l=0$ in one of the $\mathscr{E}^{(2,1)}$ factors shows that $\mathscr{E}^{(N-2,k-1)}$ is a block of $\mathscr{E}^{(N,k)}$ as well. The $N$-point conditions therefore contain lower-point ones, and relate the $N$-point correlator to the lower-point correlators.

\subsection*{Illustration with a transverse spin-J source}

It is illustrative to see how the previous constraints affect a particular case, and thus in this subsection we will consider the case of a transverse spin-$J$ in 3D, so it creates only states with $\ra{\pm J}$. Examples include a conserved current or stress energy tensor that will be studied in detail in Section~\ref{Sec:3DBounds}. 
Under a rotation by $\theta$, one has $\ra{\pm J}\to e^{\pm i J \theta}\ra{\pm J}$ and $\hat e_n\to e^{in\theta}\hat e_n$, so a matrix element requires 
\be\label{Eq:n-pointGen}
\la{h^\prime} \hat{e}_{n_1}\dots \hat{e}_{n_k} \ra{h}\,\propto\, \delta_{n_1+\dots+n_k,h^\prime-h}
\ee
so the total angular momentum of the detectors can only be 0 or $\pm 2J$. 
Because of this, and because energy and momentum conservation impose $\hat e_0=\mathbb1$ and $\hat e_{\pm1}=0$, the one-point correlator of a spin-$J$ source is described by the matrix elements
\be
\la{h^\prime} \hat{e}_n \ra{h} = H^{h^\prime h}_n \delta_{n,h^\prime-h} \,.
\ee
The only non-vanishing modes are then $n=0$ and $n=\pm 2J$. Energy conservation implies $H^{hh}_0=1$, while real energy implies $H_{2J}=H^*_{-2J}$. If one further assumes parity for simplicity, $H_{2J}=H^*_{2J}$ and is therefore real. Thus, the one-point is controlled by a single real parameter $H_{2J}$. 
The requirement of $\mathscr{E}^{(1,0)}\succeq 0$, when acted on a generic state made of $\ra{\pm J}$ states, is simply
\be
\begin{pmatrix}
    1 & H_{2J} \\ H_{2J} & 1
\end{pmatrix}
\,\succeq\,0
\ee
which implies $(H_{2J})^2\leq 1$. This is the 3D analogue of the Hofman-Maldacena bound for the one-point correlator. 

\medskip

For the two-point correlator, angular momentum conservation allows $\langle h^\prime|\hat e_n\hat e_m^\dagger|h\rangle$ to be nonzero only when $n-m=h^\prime-h$. 
This implies that $\cale(\theta_1)\cale(\theta_2)$ can be expanded in terms of a global rotation angle, or Euler angle, $\theta=\frac12(\theta_1+\theta_2)$, and a relative angle $\alpha=\theta_1-\theta_2$, as
\begin{equation}
\begin{split}
    \cale(\theta_1)\cale(\theta_2) =& \frac{1}{(2\pi)^2} \left [\quad  \sum_n \hat e_n \hat e_{-n} e^{in\alpha} \right .
    \\
    & \left . \ \ + \sum_n \hat e_{J+n} \hat e_{J-n} e^{in\alpha} e^{i2J\theta} + \sum_n \hat e_{-J+n} \hat e_{-J-n} e^{in\alpha} e^{-i2J\theta} \,+\,\dots \ \  \right ],
\end{split}
\end{equation}
where the dots represent modes whose contributions vanish in the expectation value on a transverse spin-$J$ source.
This implies that there are two families of observables for the two point functions, given by two families of multipoles. One set are the observables
\be
\la{\pm J} \hat e_n\hat e^\dagger_n \ra{\pm J} \equiv\, c_{n}\,,
\ee
corresponding to the multipole coefficients of the inclusive energy-energy correlator when expanded in terms of the relative angle $\alpha$. The other set is given by
\be
\la{ J} \hat e_{J+n}\hat e_{J-n} \ra{-J} \equiv\, d_{n}\,,\quad\quad\quad
\la{- J} \hat e_{-J+n}\hat e_{-J-n} \ra{+J} \equiv\, \bar d_{n}\,,
\ee
with $d_n=d_{-n}$ and $(\bar d_n)^*=d_n$, so $d_n$ are the multipole coefficients of the \textit{spinning} two-point correlators, and they are visible only by measuring the global position of the detectors with respect the spin's source. Energy and momentum conservation connects two-point observables with one-point ones,
\be
c_0=1\,,\quad
c_1=0\,,\quad
d_{\pm J} = H_{2J}\,,\quad
d_{\pm J\pm 1} = 0\,.
\ee
Now we can impose the positivity of two-point observables. The two conditions $\mathscr{E}^{(2,0)} \succeq 0$ and $\mathscr{E}^{(2,1)}\succeq 0$ imply constraints between the inclusive and spinning modes of the two-point correlator, for instance,
\begin{align}
\mathscr{E}^{(2,0)}\succeq 0 \quad\Rightarrow\quad\begin{pmatrix} 1 & d_n \\ d_{n}^* & 1 
\end{pmatrix}\succeq 0\,,\quad
\begin{pmatrix} 1 & c_n \\ c_{n}^* & 1
\end{pmatrix}\succeq 0\,,
\quad\text{and}\quad
\begin{pmatrix} 
1 & 0 & c_2 & c_3 & \dots \\
0 & 1 & 0 & c_2 & \\
c_2 & 0 & 1 & 0 &  \\
c_3 & c_2 & 0 & 1 & \\
\vdots & & & & \ddots
\end{pmatrix}\succeq 0\,,
\end{align}
while for the second operator, 
\begin{align}\label{Eq:3DUnitarity}
\mathscr{E}^{(2,1)}\succeq 0 \quad\Rightarrow\quad 
\begin{pmatrix} c_{n-J} & d_n \\ d_{n}^* & c_{n+J} 
\end{pmatrix}\succeq 0\,.
\end{align}
For the case of the stress energy tensor, $J=2$, we have that positivity of $\mathscr{E}^{(2,0)}$ and $\mathscr{E}^{(2,1)}$, using energy and momentum conservation, implies $0\le c_\ell \le 1$ and $|d_\ell|\leq 1$ for all $\ell$. The positive Toeplitz matrix implies nonlinear constraints among all $c_\ell$. Hermiticity of the kernel implies $(H_4)^2\leq c_4$. A measurement of $H_4$ sets a lower bound on $c_4$, with the limiting case of having $H_4$ extremal, enforcing extremality of $c_4$ as well. 
In fact, since from the Toeplitz matrix of rows $0,4,8$, one gets $c_8\to 1$ as well. This argument extends to $c_{4k}=1$ for all $k$. 
Note that higher point correlators also constrain lower ones: a measurement of $c_4$ leads to a bound on $H_4$ stronger than the one obtained from the one-point correlator alone, since it incorporates information from the two-point observables.
The remaining constraints bound the two point inclusive and spinning coefficients, $|d_n|^2\leq c_{n-J}c_{n+J}$. 

While $\mathscr{E}^{(2,0)}$ and $\mathscr{E}^{(2,1)}$ exhaust the positivity conditions stemming from two-detector operators, they do not exhaust the constraints on 2-point observables, due to energy and momentum conservation. In particular, the rows $(i,m)=(0,0),(2,2),(3,3)$ of the three point operator $\mathscr{E}^{(3,1)}$ implies
\be\label{Eq:3Dc2c3}
\begin{pmatrix}
    1 & c_2 & c_3 \\ c_2 & c_2 & 0 \\ c_3 & 0 & c_3
\end{pmatrix}
\succeq 0\,,
\ee
whose determinant is $c_2c_3(1-c_2-c_3)$. Together with its principal minors, this implies $c_2+c_3\leq1$, which is a constraint on the two-point observables independent of the two-point correlator bounds. The rest of three-point and higher-point bounds further constrain the space of inclusive and spinning observables.

\subsection*{Emergence of global positivity}

As explained in the introduction, the so-called global positivity of \eqref{Eq:GlobalPositivity} is a consequence of identifying the $N$-point correlator with a positive-semidefinite kernel, as made explicit in \eqref{eq:globalposN2} and \eqref{eq:globalposN3}. That the $N$-point correlator can be used as a positive kernel is an immediate property of representing the correlator as the average over an ensemble of events, which is made transparent with the representation of \cite{Belin:2026wkc}. 
This representation is a consequence of local positivity of $N$-point correlators for arbitrary $N$, i.e.  \eqref{Eq:LocalPositivity}, and commutativity of the energy detectors. 
This implies that global positivity is in fact an emergent property of energy-correlators in the $N\to\infty$ limit. 

\begin{figure}
    \centering
    \includegraphics[width=0.8\linewidth]{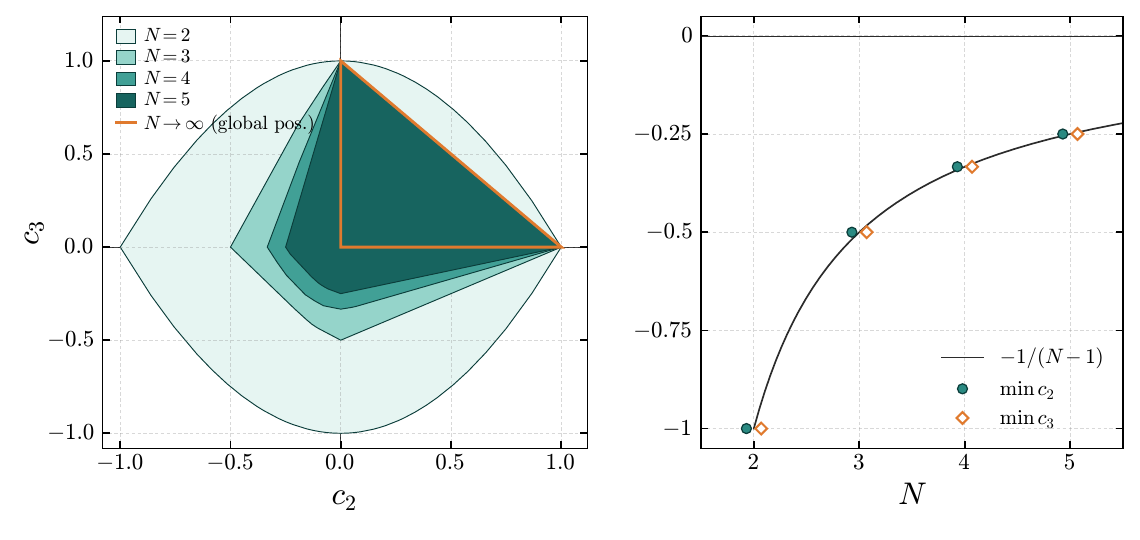}
    \caption{Illustration of how the global positivity condition emerges from local positivity conditions $\mathscr{E}^{(N,0)}\succeq 0$ in the large $N$ limit.}
    \label{fig:local_hierarchy}
\end{figure}

In order to explicitly show this point, it is interesting then to recover the bounds on the $c_2$ and $c_3$ multipoles only from the positivity of $\mathscr{E}^{(N,0)}$ at large $N$. This is shown in Fig.~\ref{fig:local_hierarchy}, where we only use $\mathscr{E}^{(N,0)}\succeq 0$ constraints in order to bound $(c_2,c_3)$. We consider $c_\ell$ up to $\ell=6$ and marginalize over the rest of observables, finding little difference by considering larger matrices. We see that $\mathscr{E}^{(N,0)}\succeq 0$ at finite $N$ allows for negative $c_2,\,c_3$, which is nonetheless ruled out by $\mathscr{E}^{(2,1)}$. However, the amount of negativity goes to zero as $1/N$, so indeed the bounds by global positivity emerge from the local bounds in the large $N$ limit. 

It is even possible to obtain the explicit lower bound in this case. Consider the function $f(\theta)=e^{i \ell \theta}$. Then, the integral $\int \prod_i^N d\theta_i |\sum_i f(\theta_i) |^2\langle \cale(\theta_1)\cdots \cale(\theta_N)\rangle$ is positive, and given by 
\be
N + N(N-1) c_\ell \geq 0 \quad\quad\quad\longrightarrow\quad\quad\quad c_\ell\geq -\frac{1}{N-1}\,,
\ee
since $\int d\theta d\theta^\prime f(\theta)f^*(\theta^\prime)\langle \cale(\theta)\cale(\theta^\prime)\rangle=c_\ell$, 
so this integral actually reproduces and saturates the range, as can be seen in the right panel of Fig.~\ref{fig:local_hierarchy}.

\subsection{Energy detectors in 4D}

The generalization of our construction to the 4D case is straightforward. Energy detectors now measure the energy in the $\vec{n}$ direction, which is now characterized by two angles. Similar to the 3D case, we decompose the energy flow in spherical harmonics,
\begin{equation}
\hat e_{\ell m}\,=\, \int d\Omega \, Y^*_{\ell m}(\vec{n})\cale(\vec{n})\,.
\end{equation}
With this definition, the energy conservation implies $\hat e_{00} = \mathbb{1}/\sqrt{4\pi}$. The momentum of massless particles is given by $\vec{p} = E \vec{n}$, so the total momentum $\int d\Omega\,\vec n\,\cale(\vec n)$ are the components corresponding to the energy operator modes of $\ell=1$. On a state at rest, total momentum conservation implies that $\hat{e}_{1m} = 0 $ for $m=-1,0,1$. Since $\cale$ is Hermitian, $\hat e^*_{\ell m}=(-1)^m\hat e_{\ell,-m}$. Under rotations, the $2l+1$ operators $\hat e_{lm}$ transform among themselves with Wigner-$D$ matrices, $\hat e_{\ell m}\to \sum_{m^\prime} D^\ell_{m m^\prime} \hat e_{\ell m^\prime}$. 

Positivity of single detector operators is imposed as in the 3D case, by ensuring that the integral over any positive function is positive. By writing $p(\vec{n})=\sum_{\ell m}c_{\ell m}Y_{\ell m}(\vec{n})$ with arbitrary $c_{\ell m}$, positivity implies
\begin{equation}
\int d\Omega\, \cale(\vec{n})|p(\vec{n})|^2\,=\,
\sum_{\ell, m, \ell^\prime, m^\prime} c^*_{\ell m} c_{\ell^\prime m^\prime} (\mathscr{E}^{(1,0)})_{\ell m;\ell^\prime m^\prime}\,\geq\, 0\,,
\end{equation}
which implies the positivity of the matrix
\begin{equation}
\label{eq:e104d}
\mathscr{E}^{(1,0)}\succeq 0\,,\quad\quad\text{with}\quad\quad
(\mathscr{E}^{(1,0)})_{\ell m; \ell^\prime m^\prime} = \sum_{L=|\ell-\ell^\prime|}^{\ell+\ell^\prime} \mathcal G^L_{\ell m ;\ell^\prime m^\prime} \,\hat e_{L,m-m^\prime}\,,
\end{equation}
where the coefficient $\mathcal G^L_{\ell m;\ell^\prime m^\prime} = \int d\Omega \, Y^*_{\ell m} Y_{\ell^\prime m^\prime} Y_{L,(m-m^\prime)}$ denotes the Gaunt integral,  namely the overlap of three spherical harmonics. 
Thus, differently from the 3D case in~\eqref{Eq:E1}, the entries of positive operators are themselves a linear combination since now different angular momentum states can couple. For a spin-one source, the positivity $\mathscr{E}^{(1,0)}\succeq 0$ in \eqref{eq:e104d} reproduces the Hofman-Maldacena bound~\cite{Hofman:2008ar} $-1/2 \leq a_\cale \leq 1$, with the one-point spinning correlator $a_\cale$ defined in \eqref{eq:onepointintro}, as illustrated in Appendix~\ref{app:sec:illustration:spinone}.

\smallskip

At two points, we can proceed as before by nesting the positivity and hermiticity conditions. 
Positivity is ensured by integrating against the square of some generic function $p(\vec{n}_1,\vec{n}_2)=\sum_{\ell_1 m_1 \ell_2 m_2}c_{\ell_1 m_1 \ell_2 m_2} Y_{\ell_1 m_1}(\vec{n}_1) Y_{\ell_2 m_2}(\vec{n}_2)$, leading to the positivity of
\begin{equation}\label{eq:4d:pov:20}
\begin{split}
\mathscr{E}^{(2,0)}\,\succeq\,0\,\quad\quad\text{with}&
\\
(\mathscr{E}^{(2,0)})_{\ell_1 m_1 \ell_2 m_2;\ell_1^\prime m_1^\prime \ell_2^\prime m_2^\prime} &= \sum_{L_1,L_2} \mathcal G^{L_1}_{\ell_1 m_1;\ell_1^\prime m_1^\prime}\mathcal G^{L_2}_{\ell_2m_2;\ell_2^\prime m_2^\prime} \,\hat e_{L_1,m_1-m_1^\prime}\hat e_{L_2,m_2-m_2^\prime}\,.
\end{split}
\end{equation}
The condition is matrix positivity in the displayed grouping of row and column indices, tested against arbitrary coefficients $c_{\ell_1m_1\ell_2m_2}$. 
We can consider the 4D version of the previous notation by denoting $A\boxtimes B$ by the arrangement of indices
\begin{align}\label{eq:squareop4d}
    \left(A\boxtimes B\right)_{\ell_1 m_1 \ell_2 m_2;\ell_1^\prime m_1^\prime \ell_2^\prime m_2^\prime} \equiv A_{\ell_1 m_1;\ell^\prime_1 m_1^\prime}\, B_{\ell_2 m_2;\ell^\prime_2 m_2^\prime}\,,
\end{align}
so that we can write $\mathscr{E}^{(2,0)} = \mathscr{E}^{(1,0)} \boxtimes \mathscr{E}^{(1,0)}$. 
Hermiticity, namely positivity of $|\sum v_{\ell m}\hat e_{\ell m}|^2$, is simply 
\begin{equation}\label{eq:4d:pov:21}
\mathscr{E}^{(2,1)} \succeq 0\,\quad\quad\text{with}\quad\quad
(\mathscr{E}^{(2,1)})_{\ell m;\ell^\prime m^\prime}\,=\, \hat e^*_{\ell m} \hat e_{\ell^\prime m^\prime}\,.
\end{equation}
In a similar manner we can construct all higher-point operators, built from the elementary operators $\mathscr{E}^{(1,0)}$ and $\mathscr{E}^{(2,1)}$ and combining them with the symbol in \eqref{eq:squareop4d}, obtaining the structure in Table~\ref{tab:ENk}.

\bigskip
Similarly to the 3D case, $ \cale(\vec{n}_1) \cale(\vec{n}_2)$ can be expanded in terms of three global rotation angles, or Euler angles, $(\Phi, \Theta, \phi)$, and the relative angle $\theta$ between two detectors.  
The energy-energy operator expanded in harmonic moments is
\begin{equation}
  \cale(\vec{n}_1) \cale(\vec{n}_2) = \sum_{\ell,m,\ell',m'} \hat{e}^*_{\ell m} \hat{e}_{\ell' m'} Y^*_{\ell m}(\vec{n}_1) Y_{\ell' m'}(\vec{n}_2)~.
\end{equation}
Two generic detectors $\vec{n}_1$, $\vec{n}_2$, while keeping internal angle $\theta$ fixed, can be obtained by acting the rotation matrix $R=R_z(\Phi) R_y (\Theta) R_z (\phi)$ on the reference detectors which are chosen to be
\begin{equation}
  \vec{n}^\text{ref}_1 = ( \sin\frac{\theta}{2}, \, 0,\, \cos\frac{\theta}{2} )~,\quad
  \vec{n}^\text{ref}_2 = ( - \sin\frac{\theta}{2}, \, 0,\, \cos\frac{\theta}{2} )~.
\end{equation} 
Accordingly, the dependence on the global angles in the product of two spherical harmonics of generic detectors is isolated into the Wigner-$D$ matrices,
\begin{equation}
  Y^*_{\ell m}(\vec{n}_1) Y_{\ell' m'}(\vec{n}_2)
  = \sum_{\lambda,\lambda'} D^\ell_{m \lambda} (R) D^{\ell' *}_{m' \lambda'} (R)  Y^*_{\ell \lambda}(\vec{n}^\text{ref}_1) Y_{\ell' \lambda'}(\vec{n}^\text{ref}_2)~,
\end{equation}
where $D^{\ell' *}_{m' \lambda'} = (-1)^{m'-\lambda'} D^{\ell'}_{-m',-\lambda'}$ and the product of two Wigner-$D$ matrices can further decompose into irreducible spin-$J$ representations with $J=|\ell - \ell'|, \cdots, \ell + \ell'$,
\begin{equation}
D^\ell_{m \lambda} (R) D^{\ell' *}_{m' \lambda'} (R) = \sum_J(-1)^{m'-\lambda'}
\langle \ell m; \ell',-m'| JM \rangle  \langle \ell \lambda; \ell',-\lambda'| JK \rangle D^J_{MK} (R)~,
\end{equation}
where $D^J_{MK} (R) = e^{-i M \Phi} d^J_{MK} (\Theta) e^{-iK\phi}$ with $M= m-m'$ and $K = \lambda - \lambda'$.
The invariance under the exchange of two detectors, which is $e^{-i K (\phi + \pi)} = e^{-i K \phi}$, allows only even $K = \lambda - \lambda'$. 
Given the selection rules, $\langle \ell m; \ell',-m'| JM \rangle \propto \delta_{M,m-m'}$ and $\langle \ell \lambda; \ell',-\lambda'| JK \rangle \propto \delta_{K,\lambda- \lambda'}$, we introduce sums over $M$, $K$ contributing only when $M= m-m'$ and $K = \lambda - \lambda'$, while varying from $-J$ to $J$. The two-point energy correlator at the operator level takes the factorized form~\cite{Riembau:2025isw},
\begin{equation}\label{eq:EE:HD}
\begin{split}
  &\cale(\vec{n}_1) \cale(\vec{n}_2) =  \sum_{J,M,K} \hat{H}^{J}_{MK}(z) D^J_{MK} (\Phi,\Theta,\phi)~,
\end{split}
\end{equation}
where $z =(1 - \vec{n}_1 \cdot \vec{n}_2 )/2$ measures the detector separation and all the inclusive and spinning correlators, denoted by $\hat{H}^{J}_{MK}(z)$, correspond to different quantum numbers $J,M,K$, and it is given by
\begin{equation}
\begin{split}
  \hat{H}^{J}_{MK}(z) &=  \sum_{\ell,m,\ell',m'} \hat{e}^*_{\ell m} \hat{e}_{\ell' m'}
   \sum_{\lambda,\lambda'} (-1)^{m'-\lambda'} 
  \\[6pt]
  &\hspace{1.5cm}\times   
\langle \ell m; \ell',-m'| JM \rangle  \langle \ell \lambda; \ell',-\lambda'| JK \rangle  Y^*_{\ell \lambda}(\vec{n}^\text{ref}_1) Y_{\ell' \lambda'}(\vec{n}^\text{ref}_2)  ~.
\end{split}
\end{equation}
For a spin-$j$ source, only values $J\leq2j$ contribute to its matrix elements.

If we are interested in the energy correlators integrated over $\Phi$, decoupling the $\Phi$-dependence or killing the interference terms between different helicities, we can consider only the matrix elements of the same helicity, $M = m - m' = h - h' = 0$:
\begin{equation}
\begin{split}
   \int_0^{2\pi}\frac{d\Phi}{2\pi}\,\cale(\vec{n}_1) \cale(\vec{n}_2) &=  \sum_{J,K} \hat{H}^{J}_{0K}(z) d^J_{0K} (\Theta) e^{-iK\phi}~.
\end{split}
\end{equation}
The inclusive correlator corresponds to the first term in the expansion with $J,M,K=0$. Using the Clebsch-Gordan (CG) coefficient
$\langle \ell m; \ell',-m'| 00 \rangle =  \delta_{\ell\ell'}\delta_{mm'} (-1)^{\ell -m}/\sqrt{2\ell + 1}$ (similarly for $\lambda$, $\lambda'$) and $D^0_{00} (R) =1$,
the inclusive term $\hat{H}^{J=0}_{M=0,K=0}(z)$ becomes
\begin{equation}
\begin{split}
\sum_{\ell=0}^\infty \sum_{m=-\ell}^\ell \hat{e}^*_{\ell m} \hat{e}_{\ell m}
 \frac{1}{2\ell + 1} \sum_{\lambda = -\ell}^\ell Y^*_{\ell \lambda}(\vec{n}^\text{ref}_1) Y_{\ell \lambda}(\vec{n}^\text{ref}_2)
 =\sum_{\ell=0}^\infty \left ( \frac{1}{4\pi}  \sum_{m=-\ell}^\ell \hat{e}^*_{\ell m} \hat{e}_{\ell m}  \right ) P_\ell (\cos\theta)~,
\end{split}
\end{equation}
where summing over $\lambda$ for the same $\ell$ makes correlator sensitive only to the internal angle between two detectors.
Defining the helicity $h$ state of the spin-$j$ source as $| h \rangle = \epsilon^{a_1\cdots a_j}_h J^{a_1\cdots a_j} |0 \rangle$, 
this defines one set of observables for the inclusive energy correlator,
\begin{equation}\label{eq:cl:incl}
 \frac{4\pi}{2\ell + 1} \sum_{m=-\ell}^\ell \langle h |   \hat{e}^*_{\ell m} \hat{e}_{\ell m}  | h \rangle \equiv c_\ell~,
\end{equation}
corresponding to multipole coefficients when expanded in the opening angle $\theta$ between two detectors. They are normalized such that $c_0=1$.
Unitarity constrains $c_\ell$ to be non-negative, $c_\ell \geq 0$~\cite{PhysRevLett.41.1581}. It can be easily shown that, beyond the inclusive energy correlator, the Legendre expansion of non-negative coefficients for spinning correlators are not possible.

Unlike the 3D case where inclusive and spinning coefficients are directly entries of the positive semidefinite matrix, the matrix elements entering the positive semidefinite matrix in 4D are not directly inclusive and spinning coefficients, denoted by $\hat{H}^J_{MK}$ in \eqref{eq:EE:HD}.
The diagonal matrix elements in source spin states, $M=m-m'=0$, are
\begin{equation}\label{eq:c:element}
    \langle h |  \hat{e}^*_{\ell m} \hat{e}_{\ell' m'} | h \rangle = \delta_{mm'} c^{(m)}_{\ell \ell'}~.
\end{equation}
The positive semidefinite condition $\mathscr{E}^{(2,1)}\succeq0$ requires the matrix to be positive semidefinite separately in each fixed-$m$ sector:
\begin{equation}\label{eq:chm:eachm}
  c^{(m)} \succeq 0,\quad \ell, \ell' \geq |m| \quad \text{for every}\ m~,
\end{equation}
with the conservation identities imposed on the relevant entries. Energy and momentum conservation give $\hat{e}_{00}=\mathbb{1}/\sqrt{4\pi}$ and $\hat{e}_{1m}=0$ for 
$m=0,\pm1$. In the $m=0$ sector, the $\ell=0$ entry is fixed as above and the $\ell=1$ row and column vanish.
In the $m=\pm1$ sectors, the $\ell=1$ rows and columns likewise vanish. The unfixed rows begin at \(\ell\geq2\) for \(m=0,\pm1\), and at \(\ell\geq |m|\) for \(|m|\geq2\); the \(m=0\) block also retains the fixed \(\ell=0\) row and its couplings.
The positive semidefinite matrix for each $m$ implies that summing over matrices of different $m$ should also be positive semidefinite, $\vec{v}^\dagger s \vec{v} \geq 0$ for any vector $\vec{v}$,
\begin{equation}\label{eq:pov:s}
s = \sum_{m}c^{(m)}   \succeq 0,\quad \ell, \ell' \geq |m|~,
\end{equation}
where the diagonal elements add up to give $c_{\ell} = 4\pi s_{\ell\ell}/(2\ell +1)$ in \eqref{eq:cl:incl}. Taking an arbitrary number of multipoles $\ell_1, \ell_2, \ell_3, \cdots$, this full positivity leads to a new constraint,
\begin{equation}\label{eq:E21matrix}
\mathscr{E}^{(2,1)}\succeq0 \quad \Rightarrow \quad
\begin{pmatrix}
   \frac{2\ell_1+1}{4\pi} c_{\ell_1} & s_{\ell_1\ell_2} & s_{\ell_1\ell_3}  & \cdots 
  \\[4pt] s^{*}_{\ell_1\ell_2} & \frac{2\ell_2+1}{4\pi} c_{\ell_2} & s_{\ell_2\ell_3}  & \cdots
  \\[4pt] s^{*}_{\ell_1\ell_3} & s^{*}_{\ell_2\ell_3} & \frac{2\ell_3 +1}{4\pi} c_{\ell_3}  & \cdots
  \\[4pt] \vdots & \vdots & \vdots 
\end{pmatrix}
 \succeq 0~,
\end{equation}
where $s_{\ell_j\ell_i} = s^{*}_{\ell_i\ell_j}$ was used. This new positivity condition, the 4D analogue of \eqref{Eq:3DUnitarity}, involves both inclusive and spinning correlators and constrains the correlations between different multipoles which is stronger than merely requiring $c_\ell \geq 0$.

The matrix elements of the positive semidefinite matrix $\mathscr{E}^{(2,0)}$ in \eqref{eq:4d:pov:20} evaluated for the same helicity source states $|h\rangle$ can be expressed in terms of elements defined in \eqref{eq:c:element},
\begin{equation}
\begin{split}
(\mathscr{E}^{(2,0)})_{\ell_1m_1\ell_2m_2;\ell_1^\prime m_1^\prime \ell_2^\prime m_2^\prime} = 
\sum_{L_1,L_2} \mathcal G^{L_1}_{\ell_1 m_1;\ell_1^\prime m_1^\prime} \mathcal G^{L_2}_{\ell_2 m_2;\ell_2^\prime m_2^\prime}  (-1)^{m_1-m_1^\prime} c^{(m_1^\prime-m_1)}_{L_1 L_2}~.
\end{split}
\end{equation}
Due to the selection rule $m_1^\prime - m_1= m_2 - m_2^\prime$, the matrix $\mathscr{E}^{(2,0)}$ should be block-diagonal in every  $m_1+m_2 = m_1^\prime + m_2^\prime$, and the positivity condition $\mathscr{E}^{(2,0)} \succeq 0$ should hold for each $(m_1+m_2)$-block. Unlike the 3D case, deriving a simple analytic form of the positive matrix is not straightforward as the matrix element appears as the weighted sum of elements of the $c$-matrix in \eqref{eq:c:element}.

The positivity condition on $c^{(m)}_{\ell \ell'}$ constrains inclusive and spinning correlators $H^{J}_{0K}(z) =\langle h| \hat{H}^{J}_{0K}(z) | h \rangle$ indirectly through the linear combination,
\begin{equation}\label{eq:HJ0K}
\begin{split}
 H^{J}_{0K}(z)  = \sum_{m} \sum_{\ell,\ell'}  c^{(m)}_{\ell \ell'} (A_{0K}^J)^{(m)}_{\ell\ell'}~,
\end{split}
\end{equation}
with the coefficients 
\begin{equation}
\begin{split}
 (A_{0K}^J)^{(m)}_{\ell\ell'} = 
   \sum_{\lambda,\lambda'} (-1)^{m-\lambda'} 
\langle \ell m; \ell',-m| J0 \rangle  \langle \ell \lambda; \ell',-\lambda'| JK \rangle  Y^*_{\ell \lambda}(\vec{n}^\text{ref}_1) Y_{\ell' \lambda'}(\vec{n}^\text{ref}_2) ~,
\end{split}
\end{equation}
where coefficients are easily evaluated and organized as an expansion in Legendre polynomials. For instance,  
for a fixed $\ell$, the coefficient $(A_{0K}^J)^{(m)}_{\ell\ell}$ includes terms up to $P_\ell(1-2z)$.
Alternatively, one could do the Legendre expansion as $H^{J}_{0K}(z) = \sum_N a^{J}_{0K,N} P_N (\cos\theta)$ and invert the relation to express $c^{(m)}_{\ell\ell'}$ in terms of $a^{J}_{0K,N}$,
\begin{equation}\label{eq:cm:in:aN}
    c^{(m)}_{\ell \ell'}
    = \sum_{J,K}\sum_{N} a^{J}_{0K,N} (B^{(m)}_{\ell\ell'})^J_{0K,N}~,
\end{equation}
where the coefficients are given by
\begin{equation}
 (B^{(m)}_{\ell\ell'})^J_{0K,N}= \int  d\mu_1d\mu_2 Y_{\ell m} ( R \vec{n}^\text{ref}_1)  P_N (\cos\theta) d^J_{0K} (\Theta) e^{-iK\phi}
     Y^*_{\ell' m} (R \vec{n}^\text{ref}_2)~,
\end{equation}
with $d\mu_1d\mu_2= d\phi d\theta \sin\theta d\Phi d\Theta \sin\Theta$.
Similarly, $c^{(m)}_{\ell \ell}$ includes up to $\ell$ number of the expansion coefficients, $N \leq \ell$.
Summing the relation in \eqref{eq:cm:in:aN} over $m$ applies to the positivity condition $s \succeq 0$ in \eqref{eq:pov:s},
\begin{equation}
    s_{\ell \ell'}
    = \sum_{J,K}\sum_{N} a^{J}_{0K,N} \sum_{m} (B^{(m)}_{\ell\ell'})^J_{0K,N}~.
\end{equation}

\section{Building positive Energy Correlators: a numerical approach}
\label{Sec:InnerNumerics}

The properties discussed in \eqref{Eq:LocalPositivity} and
\eqref{Eq:GlobalPositivity} are quite natural in theories admitting an
$S$-matrix description. Indeed, any event corresponding to a
multiparticle final state automatically satisfies these conditions.
This suggests an approach complementary to that of
Section~\ref{Sec:OuterMoments}, in which rather than deriving bounds on ECs, one
can explore their allowed region by explicitly constructing
positive energy configurations.

An event, denoted by $\mathrm{Ev}$, consists of $P$ particles, 
each carrying momentum $p_i^\mu$ and an energy fraction $x_i$ of
the total energy $E$. As before, we assume that the particles are
massless, so that their momenta can be parametrized as
$p_i^\mu=E x_i(1,\hat n_i)$, where $\hat n_i$ denotes the direction
of propagation. Energy positivity implies $ x_i\geq 0$, and energy and momentum conservation require the constraints
\begin{align}
    \sum_{i=1}^{P}x_i=1,
    \hspace{2cm}
    \sum_{i=1}^{P}x_i\hat n_i=0\,,
    \label{eq:emomcons}
\end{align}
since we work in the center-of-mass frame.

Given an event $\mathrm{Ev}=\{x_i,\hat n_i\}_{i=1}^{P}$, we can
directly compute any EC. For instance, the multipole coefficients
of the two-point EC in 4D are 
\begin{align}
    c_\ell(\mathrm{Ev})
    &=\sum_{i,j=1}^{P}
    x_i x_j P_\ell(\hat n_i\cdot\hat n_j)
    =\mathbf{x}^T C^{(\ell)}\mathbf{x}\,,
\end{align}
where
$C_{ij}^{(\ell)}=P_\ell(\hat n_i\cdot\hat n_j)$,
while one-point observables depend linearly on the energy fractions. Therefore we will generically consider an observable $\mathcal O$ evaluated on an event $\mathrm{Ev}$ by $\mathcal O(\mathrm{Ev})$. 

A collection of events
$\mathrm{Ev}_1,\ldots,\mathrm{Ev}_N$, with nonnegative weights
$w_i$ satisfying $\sum_i w_i=1$, defines a probability distribution over the space of $N$ events. The corresponding expectation value of an operator $\mathcal O$ is given by
\begin{align}
   \ab{\mathcal O}=\sum_{i=1}^{N}w_i\,\mathcal O(\mathrm{Ev}_i)\,.
\end{align}
Thus, linearity implies that any convex combination of two allowed observables, which may be evaluated on different linear set of events, is itself allowed. This immediately implies that the
region of observables generated by event distributions
is convex.

While this picture is straightforward in theories with an
$S$-matrix, Ref.~\cite{Belin:2026wkc} showed that, under suitable
positivity and consistency assumptions, ECs admit an event
distribution formula even in theories without an $S$-matrix.
Thus the event
distribution in a CFT can be represented by the same linear combination of events as presented above. Moreover, when extremizing a target observable under $K$ fixed
observables, an optimal distribution can be chosen with at most
$K+1$ events, provided the optimum is attained. The required number of particles
within each event depends on the observables being considered.
Here we propose a numerical approach to find such extremal configurations, together with the extremal values of the observables, that efficiently and systematically
constructs extremal event distributions and provides inner bounds
on the allowed region.

Consider a target observable $T$ that we want to maximize, subject
to $K$ constraints
\begin{align}
    \langle O_k\rangle=O_k^\star,
    \qquad k=1,\ldots,K.
\end{align}
Here the average implies the full set of events, meaning that $O_k^\star$ is the true value of the observable.  
Since the expectation of $T$ and $\mathcal O_k$ are linear in the event
weights, the optimization is a linear programming problem.
The difficulty is that the space of possible events is infinite.

We start by selecting a finite sample of admissible events,
$\mathcal F=\{\mathrm{Ev}_1,\ldots,\mathrm{Ev}_M\}$, and denote by
$T_{\mathcal F}^\star$ the maximum of the target observable over
all probability distributions supported on this sample.
By construction,
\begin{align}\label{eq:tfleqTstar}
    T_{\mathcal F}^\star\leq T^\star,
\end{align}
where $T^\star$ is the maximum over the full space of events.
The finite sample therefore provides an inner bound. 
Choosing $\mathcal F$ large enough so that it provides a close approximation to the continuous infinite set of events is computationally prohibitive. 
We therefore select a small set $\mathcal F$ for which the constraints admit a feasible probability distribution. 
For any observable $O$, we define their average under the ensemble $\mathcal F$ as
\begin{align}
    \langle O\rangle_{\mathcal F}
    =\sum_{\mathrm{Ev}_i\in\mathcal F}
    w_i O(\mathrm{Ev}_i).
\end{align}
The primal problem restricted to $\mathcal F$ is
\begin{align}
    T_{\mathcal F}^\star
    =\max_{\{w_i\}}\,
    &\sum_{\mathrm{Ev}_i\in\mathcal F}
    w_i T(\mathrm{Ev}_i)
    \label{eq:inner_primal}\\
    \text{subject to}\qquad
    &\sum_{\mathrm{Ev}_i\in\mathcal F}
    w_i O_k(\mathrm{Ev}_i)=O_k^\star,
    \qquad k=1,\ldots,K,
    \nonumber\\
    &\sum_{\mathrm{Ev}_i\in\mathcal F}w_i=1,
    \qquad w_i\geq0.
    \nonumber
\end{align}
The solution to this linear program leads to $T_{\mathcal F}^\star$, which obeys \eqref{eq:tfleqTstar}. 

To solve the restricted problem, we introduce $K+1$ Lagrange
multipliers $\lambda_0,\lambda_1,\ldots,\lambda_K$. 
The multiplier $\lambda_0$ enforces normalization of the event
weights, while energy and momentum conservation are already
imposed on each event through \eqref{eq:emomcons}.
The Lagrangian is
\begin{align}
    L_{\mathcal F}[w,\lambda]
    &=
    \langle T\rangle_{\mathcal F}
    +\lambda_0(1-\langle 1\rangle_{\mathcal F})
    +\sum_{k=1}^{K}\lambda_k
    \left(O_k^\star-\langle O_k\rangle_{\mathcal F}\right)
    \nonumber\\
    &=
    \lambda_0+\sum_{k=1}^{K}\lambda_k O_k^\star
    -\langle V_\lambda\rangle_{\mathcal F},
    \label{eq:inner_lagrangian}
\end{align}
where we have defined
\begin{align}
    V_\lambda(\mathrm{Ev})
    =\lambda_0+\sum_{k=1}^{K}
    \lambda_k O_k(\mathrm{Ev})-T(\mathrm{Ev}).
    \label{eq:inner_potential}
\end{align}

Maximizing the Lagrangian over the nonnegative weights gives an
unbounded result unless
$V_\lambda(\mathrm{Ev}_i)\geq0$ for every event in
$\mathcal F$. The dual problem is therefore
\begin{align}
    \text{minimize}\qquad&
    \lambda_0+\sum_{k=1}^{K}\lambda_k O_k^\star
    \label{eq:inner_dual}\\
    \text{subject to}\qquad&
    V_\lambda(\mathrm{Ev}_i)\geq0,
    \qquad \mathrm{Ev}_i\in\mathcal F.
    \nonumber
\end{align}

By linear programming strong duality, the optimal values of
\eqref{eq:inner_primal} and \eqref{eq:inner_dual} coincide:
\begin{align}
    T_{\mathcal F}^\star
    =\lambda_0^\star+
    \sum_{k=1}^{K}\lambda_k^\star O_k^\star.
    \label{eq:inner_strong_duality}
\end{align}
Furthermore, primal and dual optimality imply complementary slackness,
\begin{align}
    w_i^\star V_{\lambda^\star}(\mathrm{Ev}_i)=0,
    \qquad \mathrm{Ev}_i\in\mathcal F,
    \label{eq:inner_slackness}
\end{align}
i.e., every event with nonzero weight in the optimal distribution
saturates the dual inequality.

We now use the dual solution to improve the subset $\mathcal F$.
Although $V_{\lambda^\star}$ is nonnegative on $\mathcal F$,
it need not be nonnegative on events that do not belong to $\mathcal F$.
We therefore search for the true minimum of the potential $V_{\lambda^\star}$ on the full set (possibly infinite) of events. If the event giving this minimum is $\mathrm{Ev}_\text{min}$, given that the problem is feasible, we have that $V_{\lambda^\star}(\mathrm{Ev}_\text{min})\leq 0$. In the case of $V_{\lambda^\star}(\mathrm{Ev}_\text{min})= 0$, the extremal configuration is a convex combination of those in $\mathcal F$, and the solution $T^\star_{\mathcal F}$ is optimal, $T^\star_{\mathcal F}=T^\star$. In the case of $V_{\lambda^\star}(\mathrm{Ev}_\text{min})< 0$, the current dual certificate is violated, so we enlarge the set to $\mathcal F^\prime = \mathcal F \cup \{\mathrm{Ev}_\text{min}\}$. With this new set $\mathcal F^\prime$, we solve again the linear program of \eqref{eq:inner_primal}, in order to find a new extremal target observable $T^\star_{\mathcal F^\prime}$, such that 
\be
T^\star_{\mathcal F}\leq T^\star_{\mathcal F^\prime}\leq T^\star.
\ee
In this way we approach the target observable monotonically by incorporating the events that minimize the potential. 

The crucial step involves in finding the configuration $\mathrm{Ev}_\text{min}$ that minimizes the potential. This is done numerically by parametrizing an event in the following way. 
We choose three coordinates $\vec k_i=(k_{x,i},k_{y,i},k_{z,i})$ for $i=1,\dots,P-1$. The last vector is written as $-\sum_i \vec k_i$. The positive energy fractions are determined by $x_i = |\vec k_i|/\sum_j |\vec k_j|$. In this way, any given event $\mathrm{Ev}$ is characterized by $3(P-1)$ parameters in a way that the constraints of \eqref{eq:emomcons} are identically satisfied~\footnote{There is in fact a redundant parameter due to the global redundancy $\vec k\to t\,\vec k$. One can simply gauge fix $|\vec k_1|=1$ for a nonzero $\vec k_1$ and thus have $3P-4$ parameters. Numerically, we don't do that and we observe that this flat direction does not affect our results.}. 
Thus minimization is done by fixing the number of particles $P$. If the minimum is obtained by $p<P$ particles, the minimization itself can effectively lower the number of particles either by taking the soft limit $\vec k_i\to 0$, which implies $x_i\to 0$, or by taking the collinear limit $\vec k_i\propto \vec k_j$. In Section~\ref{Sec:results} we will observe that this is indeed the case. 
In practice, one can perform a minimization for different values of $P$ and check whether the value stabilizes. 
The minimization of the potential is a nonlinear multivariate problem and it is done numerically, and thus the obtained minimum, which may be non-optimal, $\widetilde {\mathrm{Ev}}_\text{min}$, gives only a lower bound on the true gap between the minima on $\mathcal F$ and the true global minima of the potential, i.e.
\be
T^\star_{\mathcal F}\leq T^\star_{\mathcal F\cup \{ \widetilde {\mathrm{Ev}}_\text{min} \}}\leq T^\star,\qquad -V_{\lambda^\star}(\widetilde {\mathrm{Ev}}_\text{min})\leq-\min_{\mathrm{Ev}}V_{\lambda^\star}(\mathrm{Ev}).
\ee
In the numerical implementation in Section~\ref{Sec:results}, we stop the procedure when the minimization starting from a sufficiently large number of events, gives $-V_{\lambda^\star}(\widetilde {\mathrm{Ev}}_\text{min})\leq \epsilon$, where $\epsilon$ is a given numerical precision. This gives a robust lower bound on $T^\star$, but the local stopping residual is not a certified upper bound on the remaining error. In practice, in the examples of Section~\ref{Sec:results} , $T^\star_{\mathcal F}$ converges to $T^\star$ as $\epsilon\to 0$ at least when the outer bounds of Section~\ref{Sec:OuterMoments} are known.

\medskip

For the parity-even spin-one example, with a conserved vector current $J^\mu$, in the rest frame the correlators are a matrix $H^{ij}=\la{J^i}\cale(\vec n_1)\cdots\cale(\vec n_k)\ra{J^j}$. The matrix $H^{ij}$ is positive semidefinite, which guarantees positivity for any polarized state $\epsilon^*_iH^{ij}\epsilon_j$. 
A real rank-one source weight introduces an axis $\vec s$ in the event configuration, as in \eqref{Eq:4DSourceInterval}. This breaks the $SO(3)$ degeneracy of events $\mathrm{Ev}$ mapped by a global rotation, splitting the correlator observables into the inclusive part and the spinning correlators. 

Inclusive observables, such as the multipoles $c_\ell$ of the two-point correlator, are invariant under global rotations and are evaluated on event configurations as described before, $\mathcal O(\mathrm{Ev})$. Spinning observables depend on the spin and therefore their evaluation depends on $\vec s$, $\mathcal O_{\vec s}(\mathrm{Ev})$. By rotational invariance we can fix $\vec s$ with the $\hat z$ axis for all events, so that $\mathcal O_{\vec s}$ only depends on the orientation of the event with respect $\vec s$. The simplest example is the spinning one-point correlator, $a_\cale$. Given that the full form of the correlator is given by $\ab{\cale(\vec n)}_\epsilon=\frac{1}{4\pi}\big[1+a_\cale\,(3|\epsilon\cdot\vec n|^2-1)\big]$, it can be evaluated on an event $\mathrm{Ev}$ as
\be
a_\cale(\mathrm{Ev})\,=\,\sum_i x_i\,P_2(s\cdot\hat n_i)
\ee
The algorithm is then unchanged, with events replaced with polarized events, and observables replaced by spinning observables, that carry the extra information on the spin $\vec s$. In practice, one can fix $\vec s$ along the $\hat z$ axis and let the minimization orient the event. 

\section{Applications}
\label{Sec:results}
In this section, we present some of the constraints obtained by combining the analyses of Section~\ref{Sec:OuterMoments} and Section~\ref{Sec:InnerNumerics}. 
We first consider the bounds a 3D stress-tensor source in Section~\ref{Sec:3DBounds} and then present results for a conserved spin-1 current in 4D in 
Section~\ref{Sec:4DBounds}. Finally, in Section~\ref{sec:bispecQCD}, we exemplify the bounds by constraining the bispectrum of a vector source in QCD. 

\subsection{Stress-tensor source in 3D}\label{Sec:3DBounds}

Following the discussion around \eqref{Eq:n-pointGen}, we specialize
to a source created by the stress tensor in 3D.
We use the normalized helicity states $\ket{+2}$ and $\ket{-2}$, and
write $\braket{\En{\vec{n}} \ldots }_{h'h}\equiv\bra{h'}\En{\vec{n}} \ldots \ket{h}$, with
rows and columns ordered as $(+2,-2)$. With the mode convention
$\hat e_n=\int d\theta\,e^{-in\theta}\cale(\theta)$ of
\eqref{Eq:modes3d}, the one- and two-point energy correlators take the form
\begin{align}
    \braket{\En{\vec{n}}}_{h'h}
    &=\frac{1}{2\pi}
    \begin{pmatrix}
        1 & H_4\,e^{4i\theta}\\[3pt]
        H_4^*e^{-4i\theta} & 1
    \end{pmatrix},
    \label{Eq:3DOnePointSummary}\\
    \braket{\En{\vec{n}_1}\En{\vec{n}_2}}_{h'h}
    &=\frac{1}{(2\pi)^2}
    \begin{pmatrix}
        f(\alpha)+g(\alpha) & h(\alpha)e^{4i\theta}\\[3pt]
        h(\alpha)^*e^{-4i\theta} & f(\alpha)-g(\alpha)
    \end{pmatrix},
    \label{Eq:3DTwoPointSummary}
\end{align}
where $\theta$ denotes the detector angle in the one-point function,
and $\theta=(\theta_1+\theta_2)/2$, $\alpha=\theta_1-\theta_2$
in the two-point function, as in Section~\ref{sec:3d}.
Here $f$ and $g$ are real, and all three functions are even in
$\alpha$ by detector exchange symmetry. We assume parity invariance,
so that
\begin{align}
    H_4=H_4^*,\qquad g(\alpha)=0,\qquad h(\alpha)=h(\alpha)^*.
\end{align}
Positivity of the one-point function gives $|H_4|\le1$.

\begin{figure}[t]
    \centering
    \includegraphics[width=\linewidth]{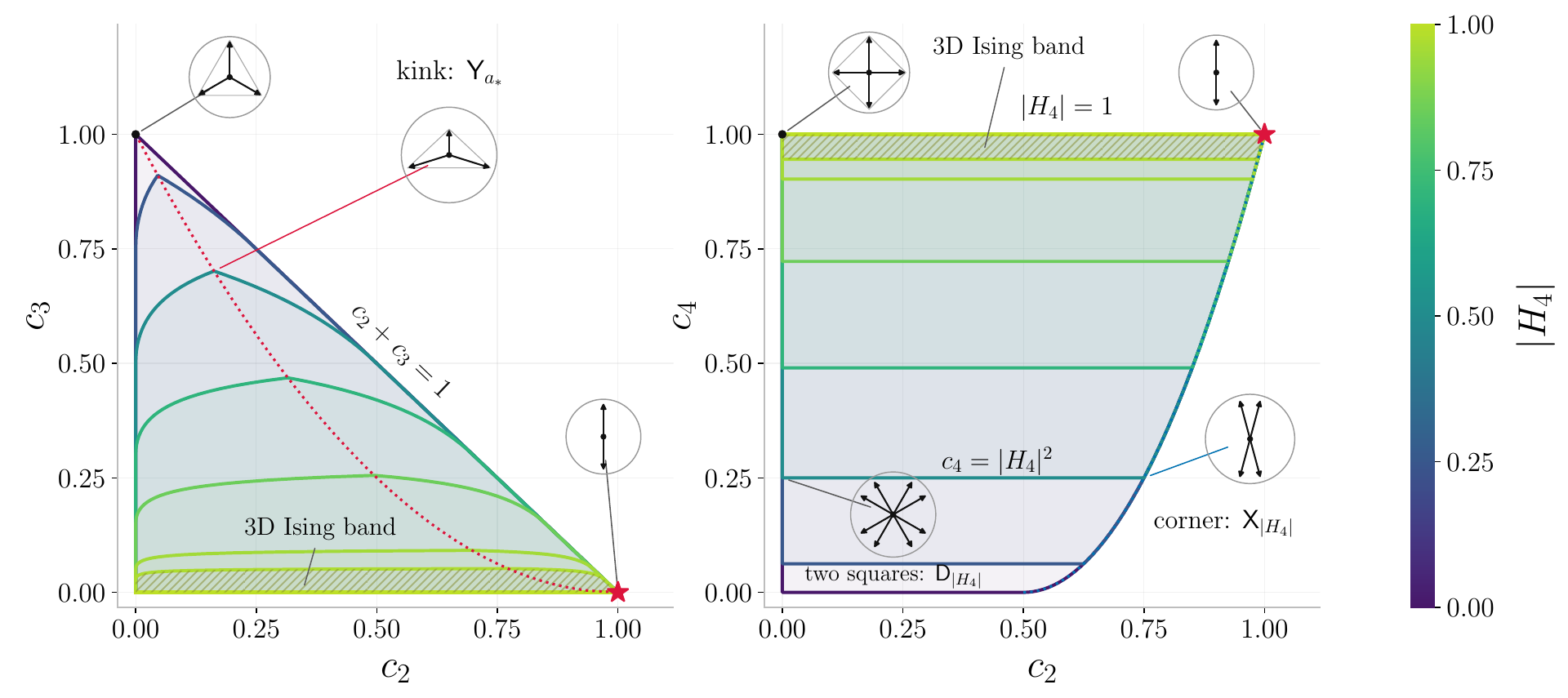}
    \includegraphics[width=0.5\linewidth]{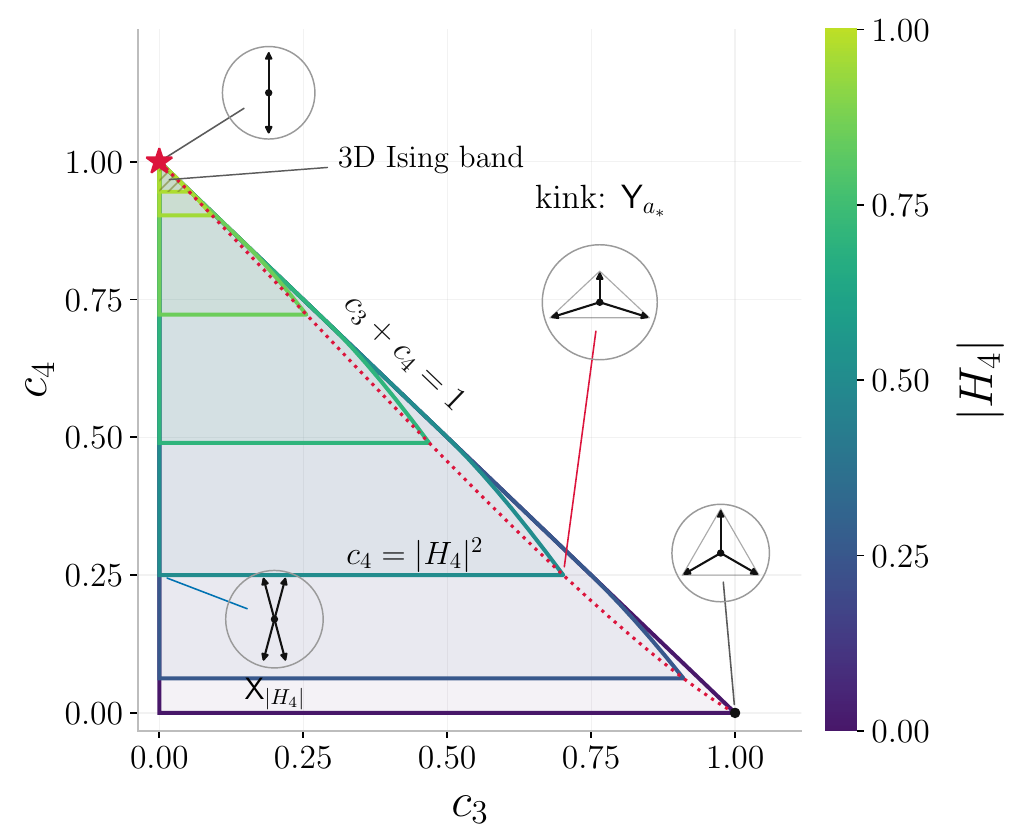}
    \caption{Allowed regions in the $(c_2,c_3)$, $(c_2,c_4)$,
    and $(c_3,c_4)$ planes for a parity-invariant spin-two source
    in 3D.
    The regions depend on the one-point parameter only through
    $|H_4|$. The highlighted Ising regions use
    $H_4=0.97233263$.
    The event configurations illustrated in the plots realize
    the corresponding boundary points.}
    \label{fig:3DRegion}
\end{figure}

We use the same inclusive and spinning multipoles as in Section~\ref{sec:3d},
\begin{align}
    c_n&=\bra{\pm2}\hat e_n\hat e_{-n}\ket{\pm2},
    &d_n&=\bra{+2}\hat e_{2+n}\hat e_{2-n}\ket{-2},\\
    f(\alpha)&=\sum_{n\in\mathbb Z}c_n e^{in\alpha},
    &h(\alpha)&=\sum_{n\in\mathbb Z}d_n e^{in\alpha}.
\end{align}
Thus $c_{-n}=c_n\in\mathbb R$ and $d_{-n}=d_n\in\mathbb R$.
Energy and momentum conservations (in the rest frame of the source) imply
\begin{align}
    c_0=1,\qquad c_{\pm1}=0,\qquad
    d_{\pm1}=d_{\pm3}=0,\qquad d_{\pm2}=H_4.
\end{align}
The plotted multipoles are $c_2,c_3,c_4$, with $0\le c_n\le1$,
and
\begin{align}
    f(\alpha)=1+2\sum_{n\ge2}c_n\cos(n\alpha).
\end{align}
In particular, \eqref{Eq:3DUnitarity} gives
$|d_n|^2\le c_{n-2}c_{n+2}$ and $H_4^2\le c_4$,
while \eqref{Eq:3Dc2c3} gives $c_2+c_3\le1$.

The allowed regions are shown in Fig.~\ref{fig:3DRegion}. Their boundaries are generated by a small
number of event families and mixtures of them (as detailed in Appendix~\ref{App:3DBounds}) and saturate the exclusion inequalities in Section~\ref{Sec:OuterMoments}, so that the bounds are effectively two-sided. The highlighted regions also show the two sided bounds for case of the Ising model, once $H_4$ is fixed to the currently known value.\footnote{More precisely $H_4$ is fixed in terms of the $\braket{TTT}$ correlators, determined by a single coefficients. The latter is usually expressed as combination of a free-boson and a free-fermion 3-point correlators: $\braket{TTT}
    =n_B\braket{TTT}_{B}+n_F\braket{TTT}_{F}$ and $H_4=\frac{n_B-n_F}{n_B+n_F}$. Table~1 of \cite{Chang:2024whx} gives
$n_B=0.933444559(75)$ and $n_F=0.013094116(33)$.} This line is merely shown to emphasize how lower point correlators put stronger constraints on higher points, but here we don't attempt to use these constraints to derive bounds on the 3D Ising CFT data which is left for future work. Details on the events saturating the boundaries and on the corresponding inequalities can be found in Appendix~\ref{App:3DBounds}.

\subsection{Conserved current source in 4D}\label{Sec:4DBounds}

We now focus on the simple case of a spin 1 source in 4D. The one-point EC is controlled by a single spinning coefficient $a_\cale$, defined as in \eqref{eq:onepointintro}. As explained in Section~\ref{Sec:intro}, extremal values of $a_\cale$ constrain higher points by forcing the underlying events to be collinear or planar. 
The bounds of Section~\ref{Sec:OuterMoments} and Section~\ref{Sec:InnerNumerics} allow us to quantify the impact on higher-point correlators of a non-extremal value of $a_\cale$. 

\begin{figure}[t]
    \centering
    \includegraphics[width=\linewidth]{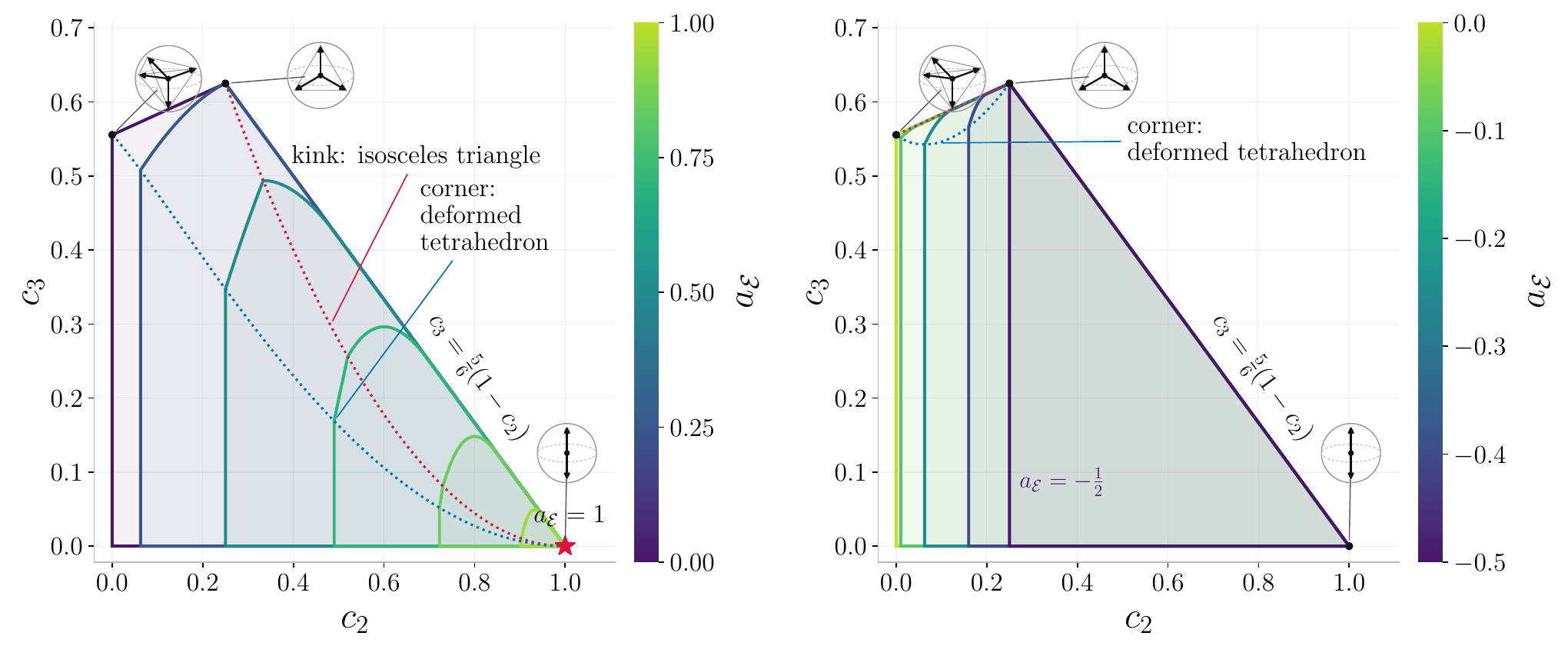}
    \caption{Spin-one event regions in the $(c_2,c_3)$ plane.
    Left: $a_\cale=0,0.25,0.5,0.7,0.85,0.95$.
    Right: $a_\cale=-0.5,-0.4,-0.25,-0.1,0$, with a separate
    color scale for $-1/2\leq a_\cale\leq0$. We use
    $c_\ell=\sum_{ij}x_i x_j P_\ell(\vec n_i\cdot\vec n_j)$,
    with $c_0=1$.
    The $a_\cale=0$ region coincides with the inclusive one, and
    $a_\cale=1$ gives the back-to-back event (star).
    The curves are analytic atomic-event constructions.
    The sketches indicate atomic events realizing the marked boundary
    points, with schematic ray lengths proportional to energy fractions.
    The blue dotted curves trace the deformed-tetrahedron corners in both
    panels. The red dotted curves trace the isosceles-triangle kinks on the
    left and, on the right, the points where mixtures of equilateral
    triangles and deformed tetrahedra meet the straight upper boundary.
    See Appendix~\ref{App:4D23} for the analytic boundaries and their
    saturating events.}
    \label{fig:4DRegionsAnalytic}
\end{figure}

The allowed region of the multipole coefficients $c_2$ and $c_3$ of the inclusive part of the two-point energy correlator, once the one-point spinning $a_\cale$ is fixed, are shown in Fig.~\ref{fig:4DRegionsAnalytic}. 
In terms of energy fractions $x_i$, these are defined and normalized as $c_\ell = \sum_{ij}x_i x_j P_\ell(\vec n_i\cdot \vec n_j)$, so that $c_0=1$ and two back-to-back particles give $c_\ell=1$ for even $\ell$. 
With a single constraint, the boundaries of parameter space can be worked out analytically like in the 3D case. 
For $a_\cale=0$, namely a source compatible with a scalar source, the allowed parameter space in the $(c_2,c_3)$ plane coincides with the inclusive case, as in \cite{Belin:2026wkc}. The point at the origin can be obtained either by a complete isotropic distribution or by a simple multiparticle event with enough particles to get the $c_i = 0$. The vertices are given by a two-, three- and four-particle event. For positive $a_\cale$, the allowed region shrinks to zero as $a_\cale\to 1$, with the minimum of $c_2$ given by $\, a_\cale^2$. The regions for finite $a_\cale$ have two kinks, corresponding to where the boundary is given by a single event, corresponding to an isosceles triangle or a deformed tetrahedron, which descend from the equilateral and tetrahedral events at the vertices of the boundary of the inclusive case, respectively. For negative $a_\cale$, the situation is qualitatively different as the extremal point $a_\cale=-\frac12$ is compatible with different configurations. The analytic boundaries and their saturating events are given in Appendix~\ref{App:4D23}.

\begin{figure}
    \centering
    \includegraphics[width=\linewidth]{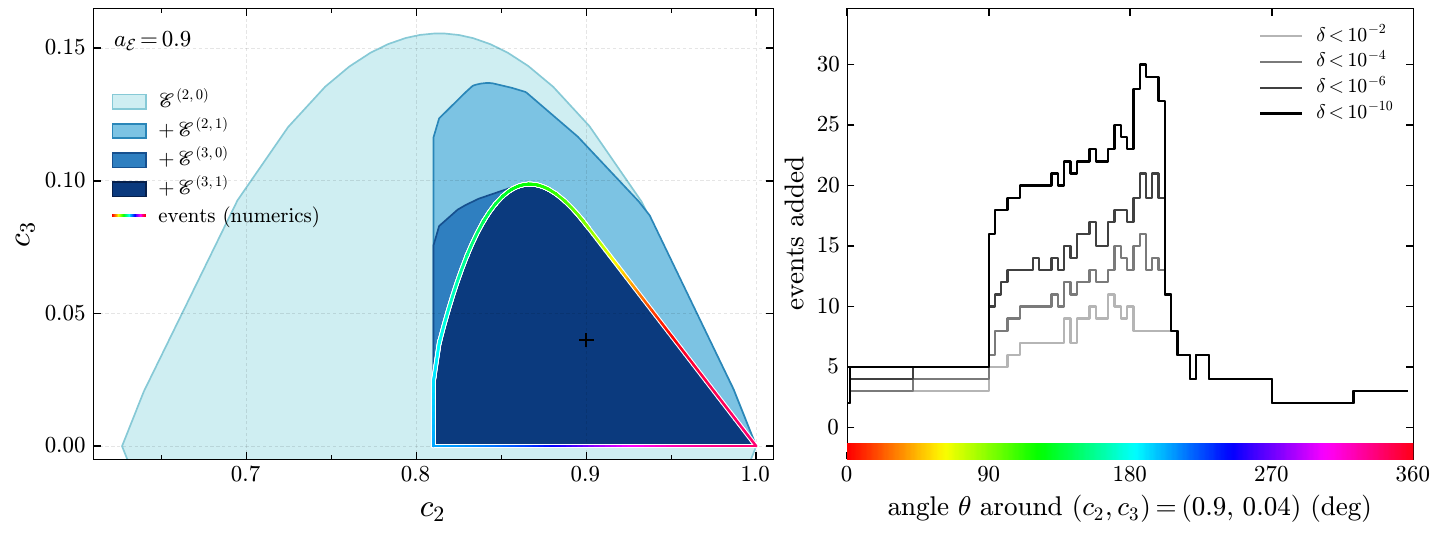}
    \caption{Left: Constraints on the $c_2$ and $c_3$ observables from a spin 1 source fixing the spinning one-point coefficient to $a_\cale=0.9$. Different shades of blue correspond to the allowed regions after imposing the positivity of the $\mathscr{E}^{(N,k)}$ operators. Colorful line corresponds to the space reachable by the numerical approach. Right: events added in the numerical minimization, together with the obtained local separation residual $\delta=-V_{\lambda^\star}(\widetilde {\mathrm{Ev}}_\text{min.})$.}
    \label{fig:fig_a09}
\end{figure}

In order to illustrate how the different bounds in Section~\ref{Sec:OuterMoments} and Section~\ref{Sec:InnerNumerics} converge in this simple case, we choose $a_\cale=0.9$ and compare the different constraints in Fig.~\ref{fig:fig_a09}. 
The regions in different shades of blue show the allowed regions once imposing the positivity of the $\mathscr{E}^{(N,k)}$ operators. Since these operators give the outer bounds, the allowed region shrinks when adding higher-point detector operators. 
Positivity of $\mathscr{E}^{(2,0)}$ alone already reduces significantly the allowed region in the $(c_2,c_3)$ plane compared with the inclusive case. Global positivity, namely $\mathscr{E}^{(2,1)}\succeq 0$, further reduces the region. Since the norm of the states $\hat e_{\ell m}\ra{h}$ is given by $c_\ell$, $\sum_m ||\hat e_{\ell m}|h\rangle ||^2=\frac{2\ell+1}{4\pi}c_\ell$,  this implies $c_\ell\geq0$ for every $\ell$. 
Positivity of $\mathscr{E}^{(2,1)}\succeq 0$ further implies a relation between the two-point multipoles and the one-point spinning parameter, $a_\cale^2\leq c_2$, stemming from \eqref{eq:E21matrix}. This implies the vertical edge at $c_2=0.81$. 
This bound is seen also at the event level in \eqref{Eq:4DSourceInterval}, which gives $a_\cale(\mathrm{Ev})^2\leq c_2(\mathrm{Ev})$. Positivity of three detector operators removes the upper part of the region. The $\mathscr{E}^{(3,0)}$ contains the inequality $5c_2+6c_3\leq5$,\footnote{For
$\vec V=\sum_{\mathrm{cyc}}(\vec n_1\cdot\vec n_2)(\vec n_1-\vec n_2)$
and $D=\vec n_1\cdot(\vec n_2\times\vec n_3)$, we have
\begin{align}
0\leq\int\prod_{i=1}^3d\Omega_i\,
\Big\langle\prod_{i=1}^3\cale(\vec n_i)\Big\rangle
\bigl(|\vec V|^2+3D^2\bigr)
=\frac25(5-5c_2-6c_3).
\end{align}
} which gives the flat region and is saturated by a convex combination of the back-to-back event at $(c_2,c_3)=(1,0)$ and three-particle events. The $\mathscr{E}^{(3,1)}$ bounds agree with the inner region at the plotted numerical resolution. 

The inner region, shown as a colorful boundary in Fig.~\ref{fig:fig_a09}, is obtained with the algorithm presented in Section~\ref{Sec:InnerNumerics} and converges to machine precision with the analytic boundaries (see Fig.~\ref{fig:4DRegionsAnalytic}). We parametrize the $(c_2,c_3)$ plane with polar coordinates centered at $(0.9,0.04)$, and extremize the radius $r$ for a grid of 90 values of the angle once $a_\cale$ is fixed to $a_\cale=0.9$. A finer resolution to resolve the $(1,0)$ vertex is added. 
The algorithm starts with a pair of back-to-back 2-particle events, and extra events are added by numerically minimizing the potential. 
The right of Fig.~\ref{fig:fig_a09} shows the events added to the initial configuration via the minimization as a function of the angle $\theta$ in the $(0.9,0.04)$-centered coordinates, together with the reached gap between the primal solution $T^\star_{\mathcal F}$ in the given set and the negativity of the potential obtained via the numerical minimization, given by $-\delta=V_{\lambda^\star}(\widetilde {\mathrm{Ev}}_\text{min.})$. We observe that in the flat boundaries convergence at the $\delta<10^{-10}$ level is obtained already with a handful of events. Instead, in the curved boundary order $20$ events should be added in order for the algorithm to reach $\delta<10^{-10}$ convergence. The kink between the $c_2=0.81$ flat constraints and the one coming from the $\mathscr{E}^{(3,1)}$ shows the slowest convergence. 

In terms of computing time, the numerical approach for the inner bounds is significantly faster than the outer bounds, even asking for $\delta<10^{-10}$ convergence. Extremizing an observable along a single ray in a flat edge takes a few seconds on a single core, and about two minutes at most to reach $\delta<10^{-11}$ in the worst case near the kink. For comparison, the outer bounds, for the semidefinite minimization in the same core,  cost around a minute for $\mathscr{E}^{(3,0)}$ and about 10 minutes for $\mathscr{E}^{(3,1)}$. 

\medskip

\subsection{Bispectrum constraints in QCD}
\label{sec:bispecQCD}

As a physical application we consider $e^+e^-\to\text{hadrons}$ in the $Z$-pole, where the source is the spin-1 current the $Z$-boson couples to. 
We impose the measured one- and two-point correlators. The one-point correlator was extracted in \cite{Riembau:2025wjc}, 
reinterpreting DELPHI and OPAL measurements, giving  $a_\cale=-0.423\pm 0.011$ \cite{DELPHI:1997oih} and $a_\cale=-0.415\pm 0.008$ \cite{OPAL:1995xgn}, respectively. We consider the weighted average, $a_\cale=-0.418$. For the two-point correlator, the multipoles $c_\ell$ for $\ell\leq 20$ were measured in \cite{Jaarsma:2025tck,Electron-PositronAlliance:2025fhk}. We consider the mean value to fix the multipoles.\footnote{In particular, we fix the multipoles to the central values of the data points of Fig.~21 of Ref.~\cite{Jaarsma:2025tck}, $c_\ell=0.742, 0.599, 0.496, 0.419, 0.360, 0.315, 0.279, 0.251, 0.229, 0.211$ for even $\ell=2,\dots,20$, and $c_\ell=0.092, 0.117, 0.132, 0.140, 0.145, 0.147, 0.147, 0.147, 0.147$ for odd $\ell=3,\dots,19$, but the uncertainties are estimated to be around $\pm 0.02$ for all of them. The measurement is done on charged tracks and the uncertainty of the measurement is reported in \cite{Jaarsma:2025tck} as only an estimate.}

\begin{figure}
    \centering
    \includegraphics[width=0.7\linewidth]{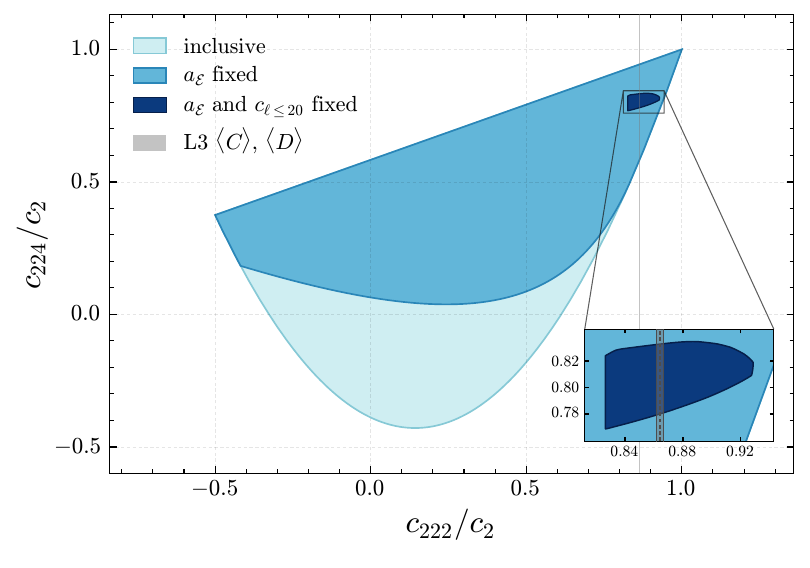}
    \caption{Allowed region in the bispectrum observables $c_{222}$ and $c_{224}$ normalized to the two-point $c_2$. In light blue, inclusive region. In blue, allowed region after assuming a spin 1 source with $a_\cale$ fixed to the observed QCD value. In dark blue, the numerical inner region obtained after fixing both the one-point $a_\cale$ and the two-point observables $c_\ell$ for $\ell\leq 20$ to the observed QCD values. The vertical grey band shows the inferred value from the L3 measurement of the $C$ and $D$ parameters.}
    \label{fig:bispec_fig}
\end{figure}

With this input, we constrain observables related to the three-point correlator. 
At the inclusive level, a basis of observables, in analogy with CMB literature \cite{Komatsu:2001rj}, are the (reduced) bispectrum coefficients
\bea\nonumber
c_{\ell_1\ell_2\ell_3} &=& \begin{pmatrix}   \ell_1 & \ell_2 & \ell_3 \\ 0 & 0 & 0 \end{pmatrix}^{-2}
 \int \frac{d\Omega}{4\pi} \int \prod_{i=1}^3 d\Omega_i P_{\ell_i}(\vec{n}_i\cdot \vec{n})\times \langle \cale(\vec{n}_1)\cale(\vec{n}_2)\cale(\vec{n}_3)\rangle\,\\
&=&
\mathcal{N}_{\ell_1\ell_2\ell_3}\sum_{m_1,m_2} \mathcal G^{\ell_3}_{\ell_1 m_1;\ell_2 m_2} \langle \hat e^*_{\ell_1 m_1}\hat e_{\ell_2 m_2}\hat e_{\ell_3 ,m_1-m_2}\rangle\,,
\label{eq:bispectrumdef}
\eea
Here the multipoles obey the triangle inequalities and $\ell_1+\ell_2+\ell_3$ is even, so the Wigner symbol in the denominator is nonzero. The coefficients $\mathcal G^{\ell_3}_{\ell_1 m_1;\ell_2 m_2}$ are the integrals in \eqref{eq:e104d} and $\mathcal{N}_{\ell_1\ell_2\ell_3}=(4\pi)^2(\ell_1 \ell_2 \ell_3; 0  0  0 )^{-2}((2\ell_1+1)(2\ell_2+1)(2\ell_3+1))^{-1}$. 
The second line follows from $P_\ell(\vec n\cdot \vec n^\prime) = \frac{4\pi}{2\ell+1}\sum_m Y^*_{\ell m}(\vec n) Y_{\ell m}(\vec n^\prime)$. These are the generalization of the multipole observables $c_\ell=c_{\ell\ell 0}$ at the two-point level. Due to energy and momentum conservation, the first genuine three-point observables are $c_{222}$ and $c_{224}$. 

In Fig.~\ref{fig:bispec_fig} we show the allowed  $(c_{222},c_{224})$ plane, normalized by $c_2$. This light blue region coincides with \cite{Belin:2026wkc} and refer to the inclusive case, without additional constraints. In particular, $-\frac12 \leq c_{222}/c_2\leq 1$. In blue, we show the allowed region after we fix the spinning one-point observable $a_\cale$ to its observed average value of $a_\cale=-0.418$.\footnote{The analytic bounds for the inclusive and fixed-$a_\cale$ bands, together with their saturating events, are given in Appendix~\ref{App:4DBispectrum}, \eqref{Eq:4DBispectrumInclusive} and~\eqref{Eq:4DBispectrumLower}.} In dark blue, we show a numerical inner region after additionally fixing the $c_\ell$ multipoles with $\ell\leq20$. We follow the algorithm of Section~\ref{Sec:InnerNumerics}, starting with a large number of events and asking convergences of at least $\delta<10^{-3}$ in the whole boundary. 
We see that the two-point observables, measured at the QCD value, set strong constraints on the three point ones. 
The difference between the one- and two-point ECs is that the first, via $a_\cale\simeq -\frac12$, constrains event planarity without fixing the particle content; the extremal value is compatible both with two free fermions minimally coupled to the vector and with the WZW current. It is the measurement of the two-point correlator what discriminates these degeneracies and significantly constrains the three-point observables. 

To the best of our knowledge, there is no full direct measurement nor computation of the three-point correlators and therefore no prediction for the $c_{\ell_1\ell_2\ell_3}$ in QCD apart from $c_{222}$, as we shall discuss.\footnote{A tree-level computation of the three-point correlator at generic angles was performed in \cite{Yang:2022tgm}. However, its integral over the full angular phase space diverges at the planar boundary, where the one-loop three-parton contributions must be included, together with the contact terms. The coplanar limit of the three-point energy correlator was subsequently studied in \cite{Gao:2024wcg}. It does not, however, provide the complete inclusive three-point correlator needed to evaluate the full angular integral.} However, the $c_{222}$ coefficient can be mapped to the well-known $C$- and $D$-parameters \cite{Parisi:1978eg,Donoghue:1979vi,Ellis:1980wv}, measured at LEP \cite{L3:2004cdh,OPAL:2004wof}. These are defined by considering the momentum tensor $\theta^{ab}=\frac{1}{\sum_j |\vec p_j|}\sum_i \frac{p_i^a p_i^b}{|\vec p_i|}$.\footnote{Here, as well as in the procedure used to obtain the bounds, corrections proportional to the hadron masses are neglected.} Normalizing $\theta$ to have unit trace, the $C$- and $D$-parameters are defined by the two symmetric combinations of its eigenvalues $C=3(\lambda_1\lambda_2+\lambda_2\lambda_3+\lambda_3\lambda_1)$ and $D=27\lambda_1\lambda_2\lambda_3$, see \cite{Ellis:1980wv}. 
In the massless approximation used here, the momentum tensor can be written as $\theta^{ab}=\int d \Omega_{\vec{n}} n^a n^b \cale(\vec{n})$, related to the traceless symmetric part of the one-point correlator, proportional to $A^{ab}\equiv3 \theta^{ab}-\delta^{ab}$. Therefore, the five independent entries correspond to the five components $\hat e_{2m}$. Thus, there is a relation between the momentum tensor and the one-point correlator by $\int d\Omega_{\vec{n}'}P_2(\vec{n}\cdot \vec{n}^\prime)\cale(\vec{n}^\prime) = \frac{4\pi}{5}\sum_m Y_{2m}(\vec{n})\hat e_{2m} = \frac12 \vec{n}\cdot A\cdot \vec{n}$. Therefore, correlators of powers of $\int d\Omega_{\vec{n}'} P_2(\vec{n}\cdot \vec{n}^\prime)\cale(\vec{n}^\prime)$, which corresponds to the $c_2$ multipole at two-points and the $c_{222}$ bispectrum at three points, see \eqref{eq:bispectrumdef}, corresponds to the second and third power of this operator, which are related to the products of eigenvalues $C$ and $D$ by $C = 1-\tfrac16\,\mathrm{tr}A^2$ and $D = 1-\tfrac12\,\mathrm{tr}A^2+\tfrac13\,\mathrm{tr}A^3$, which are operator identities and thus hold event-by-event. On the other hand, $c_2=\tfrac16\langle\mathrm{tr}A^2\rangle$ and $c_{222}=\tfrac16\langle\mathrm{tr}A^3\rangle$, leading to
\be
\ab{C} = 1-c_2\,\quad\quad\quad\text{and}\quad\quad\quad
\ab{D} = 1-3c_2+2 c_{222}\,.
\ee
Thus the second relation shows that it is possible to compare the bounds on $c_{222}$ to the LEP measurements of $\ab{C}$ and $\ab{D}$. L3 results in \cite{L3:2004cdh}, Tables 57 and 58, give the $C$-parameter separately for $udcs$ quarks and for $b$-quarks. The weighted average gives $\ab{C}=0.2640\pm 0.0021$, 
and $\ab{D}=0.0642\pm 0.0010$. This is compatible with OPAL measurement of $\ab{C}=0.2646\pm 0.0023$ in \cite{OPAL:2004wof}, and compatible with their reported measurement of the $\ab{D}$ distribution in Table~6 and Fig.~10 of \cite{OPAL:2004wof}, whose integral, assuming a flat distribution in each bin gives $\ab{D}\simeq 0.065$. 
Together these give 
\be
\frac{c_{222}}{c_2} \,=\, 1-\frac12\frac{\ab{C}-\ab{D}}{1-\ab{C}} \,=\,0.864\pm 0.003\,,
\ee
as shown in the grey vertical band in Fig.~\ref{fig:bispec_fig}, which indeed lies well inside the region indicated by positivity. 
In contrast, $c_{224}$ involves the modes $\hat e_{4m}$ and it is a genuinely new three-point observable, for which we only have the positivity bounds.

\section{Conclusions}

Energy correlators are elementary observables in scattering experiments. Positivity of the energy, in the form of \eqref{Eq:LocalPositivity} and \eqref{Eq:GlobalPositivity}, implies an infinite set of constraints on the possible functional forms of multipoint ECs. 
This work provides the complete set of constraints by finding, in a simple and systematic way,  a classification of all positive operators that can be built from $N$-detectors, which we call $\mathscr{E}^{(N,k)}$, and follow the simple construction shown in Tab.~\ref{tab:ENk}. While one needs to truncate the problem at some $N$ and $k$, the positive operators provide necessary conditions for an observable to be admissible, providing robust \textit{outer} bounds. 

Furthermore, in Section~\ref{Sec:InnerNumerics} we provide a simple linear program in order to find the extremal observables compatible with a given set of constraints. This is obtained by considering a finite set of energy configurations that can be extended systematically by a numerical minimization of a potential in event space. This approach constructs robust \textit{inner} bounds for the observables, with the boundary given by explicit energy configurations. 

Combining both we are able to obtain two sided bounds on ECs, both for inclusive and spinning sources, and also connecting lower and higher point ECs. Results in Section~\ref{Sec:results} show instances of these findings, both in 3D and 4D cases. For instance, in Fig.~\ref{fig:bispec_fig} we show how the spinning one-point correlator and inclusive two-point observables in QCD strongly constrain the bispectrum.  One of these multipoles, $c_{222}$, is given by LEP measurements of the $C$- and $D$-parameters, which fall well inside the predicted island, while the construction explores attainable values of $c_{224}$, a three-point observable that has not been measured directly.  

An important and sometimes underappreciated point is that when there is a representation of the correlators in terms of events, positivity constraints on ECs are an obvious and trivial consequence of rotational invariance and energy positivity, as the numerical approach in Section~\ref{Sec:InnerNumerics} explicitly exploits. In particular, any showering or hadronization algorithm producing positive-energy, momentum-conserving events will satisfy these constraints. Nevertheless, the constraints may have nontrivial implications for theoretical and experimental systematics. Furthermore, positivity constraints can help characterize the allowed shapes of energy correlators in exotic jets arising from New Physics. This is particularly useful both in weakly coupled scenarios, where a description in terms of asymptotic particles is available but one wishes to remain more or less agnostic about the underlying model \cite{Ricci:2026fkj}, and in strongly coupled scenarios, where detailed predictions for final states may be difficult to obtain, while inclusive observables such as ECs can remain comparatively tractable \cite{Ricci:2026qpw}.

Positivity of ECs can play an important role whenever the correlators do not have an obvious representation in terms of events. This is, indeed, the case for CFT where they imply nontrivial constraints on CFT data, see e.g. \cite{Hofman:2008ar,Cordova:2017zej,Meltzer:2018tnm,Mecaj:2025ecl,Dempsey:2025yiv,Belin:2026wkc}. For instance, an interesting application might be to see the implications of the ECs positivity bounds of Section~\ref{Sec:3DBounds} to the 3D Ising model, where the precisely known $\braket{TTT}$ correlator coefficient can imply strong constraints on CFT data entering $\braket{TTTT}$. 
In general, the obtained constraints provide the basis toward deriving the full set of constraints on CFT data involving the stress energy tensor due to energy positivity. A precise assessment of these implication is left for future work.

Although we have focused on energy detectors, our framework could be extended to a broader class of positive detectors. In particular, \cite{Kravchuk:2018htv,Meltzer:2018tnm} established a higher-spin generalization of ANEC, showing that, under suitable assumptions, the appropriately normalized leading-twist, even-spin detectors are positive semidefinite operators. In free theories, such operators correspond to measuring powers of the energies of individual particles, making it possible to perform an inner construction analogous to that of Section~\ref{Sec:InnerNumerics}. By contrast, an outer construction along the lines of Section~\ref{Sec:OuterMoments}, the existence of a corresponding event distribution formula, and the convergence of the inner and outer bounds generally require additional assumptions. We leave these generalizations for future work.

\paragraph{Acknowledgments}

We thank J.~Henriksson and A.~Zhiboedov for discussions and comments on the draft. 
MR is supported by the Swiss National Science Foundation under contract 200020-213104. 
MS was supported by the KAIST Strategic Research Program (N10260105, N10260106). The authors acknowledge the use of OpenAI Codex and Anthropic Claude Code for feedback on the manuscript, development of numerical routines and cross-check of analytic results. The authors take full responsibility for the manuscript.

\appendix

\section{Stress-tensor source in 3D}\label{App:3DBounds}

\subsection{Event basis}

The boundaries in Fig.~\ref{fig:3DRegion} can be simply described by the superposition of at most two events (denoted $E_i$) with positive weight adding up to 1. The basis of events we need is 
given below.

We take the reference axis and the source-polarization axis along the
vertical $\theta =0$ direction. Single-direction angles are measured counterclockwise
from this axis; the overall orientation of an event remains arbitrary.
An $N$-particle event is specified by its reference angles $\beta_i$ on the
circle and energy fractions $x_i\ge0$, satisfying
\begin{align}
    \sum_i x_i=1,\qquad \sum_i x_i e^{i\beta_i}=0.
\end{align}
After averaging over the overall orientation, its inclusive correlator is
\begin{align}
    \frac{f(\alpha)}{2\pi}
    &=\sum_{i,j}x_ix_j\,
      \delta_{2\pi}\!\left(\alpha-(\beta_i-\beta_j)\right),
    \label{Eq:3DEventPairWeights}
\end{align}
Here $\delta_{2\pi}$ is the periodic delta function with unit integral.
The sum runs over ordered pairs, including the self-pairs $i=j$.
Thus each peak carries the sum of the corresponding energy products
$x_ix_j$; only for equal energies are its weights purely combinatorial.
We retain the oriented angle $\alpha=\theta_1-\theta_2$ modulo $2\pi$,
so the two signs of a separation appear separately.
The drawings below label every particle by its reference angle $\beta$
and energy fraction $x$; the rays indicate directions, not energy magnitudes.

The spinning function $h$ is obtained similarly. With the polarization axis
at $\theta=0$, rotate the whole event by $\psi$:
$\beta_i\rightarrow\beta_i+\psi$. For the reflection-symmetric events below,
assign the off-diagonal weight $q e^{4i\psi}$ to the event, with real
$q$ and $|q|\le1$. Then
\begin{align}
    \frac{e^{4i\theta}h(\alpha)}{(2\pi)^2}
    &=\bra{+2}\En{\vec{n}_1}\En{\vec{n}_2}\ket{-2}\nonumber\\
    &=q\sum_{i,j}x_ix_j\int_0^{2\pi}\frac{d\psi}{2\pi}\,e^{4i\psi}
      \delta_{2\pi}(\theta_1-\beta_i-\psi)
      \delta_{2\pi}(\theta_2-\beta_j-\psi),
    \label{Eq:extractingh}
\end{align}
where $\theta=(\theta_1+\theta_2)/2$ and $\alpha=\theta_1-\theta_2$.

Applying the same average to the one-point function gives
\begin{align}
    H_4=q\,e_4,\qquad e_4=\sum_i x_i e^{-4i\beta_i},
    \qquad |H_4|^2\le |e_4|^2=c_4.
    \label{Eq:EventH4}
\end{align}
For a fixed geometry with $e_4\ne0$, this determines $q=H_4/e_4$.
If $e_4=0$, then $H_4=0$ but $q$ is not fixed by the one-point function.

\begin{itemize}
    \item \textbf{Uniform energy flow.}
    \begin{align}
        \text{E}_0: && \braket{\En{\vec{n}_1}\En{\vec{n}_2}}_{h'h}
        =\frac{\delta_{h'h}}{(2\pi)^2}\,, && \braket{\En{\vec{n}}}_{h'h}= \frac{\delta_{h'h}}{2\pi}\,,
    \end{align}
    which gives
    \begin{align}
        (c_2,c_3,c_4)=(0,0,0),
        \qquad H_4=0.
    \end{align}
    Equally spaced events with $N\ge5$ particles, each of energy $1/N$, have
    the same values of $c_2,c_3,c_4$ and $H_4$.
    \item \textbf{Back-to-back event $\mathsf B$.}
    Two particles with energies $1/2$ lie at reference angles $0$ and $\pi$.
    The drawing shows a rotated copy at angles $\psi$ and $\psi+\pi$
    relative to the fixed $z$ axis.
    \nopagebreak[4]
    \input{figs/events/back_to_back.tex}
    Here $e_4=1$, so $q=H_4$. The correlator is
    \begin{align}
        E_{\mathsf{B}}: && \braket{\En{\vec{n}_1}\En{\vec{n}_2}}_{h'h}
        =\frac{\delta_{2\pi}(\alpha)+\delta_{2\pi}(\alpha-\pi)}{4\pi}
        \begin{pmatrix}
            1&H_4 e^{4i\theta_1}\\
            H_4 e^{-4i\theta_1}&1
        \end{pmatrix}.
    \end{align}
    Thus
    \begin{align}
        (c_2,c_3,c_4)=(1,0,1),\qquad |H_4|\le1.
    \end{align}

    \item \textbf{Equilateral triangle $\mathsf T$.}
    Three particles with energies $1/3$ at reference angles
    $0,2\pi/3,4\pi/3$ form an equilateral triangle.
    \nopagebreak[4]
    \input{figs/events/triangle.tex}
    Their correlator is
    \begin{align}
        E_{\mathsf T}: && \braket{\En{\vec{n}_1}\En{\vec{n}_2}}_{h'h}
        =\frac{\delta_{h'h}}{6\pi}\sum_{k=0}^{2}
        \delta_{2\pi}\!\left(\alpha-\frac{2\pi k}{3}\right),
    \end{align}
    and hence
    \begin{align}
        (c_2,c_3,c_4)=(0,1,0),\qquad H_4=0.
    \end{align}
    In \eqref{Eq:extractingh}, the three phases sum to
    $1+e^{-8\pi i/3}+e^{-16\pi i/3}=0$, so the off-diagonal part vanishes
    for every $q$. In particular, $e_4=H_4=0$.
    \item \textbf{Square event $\mathsf Q$.}
    Four particles with energies $1/4$ lie at reference angles $k\pi/2$,
    $k=0,1,2,3$. Here $e^{-4i\beta_i}=1$ for every particle, so $e_4=1$
    and $q=H_4$: the off-diagonal part does not cancel in the orientation average.
    \nopagebreak[4]
    \input{figs/events/square.tex}
    Their correlator is
    \begin{align}
        E_{\mathsf Q}: && \braket{\En{\vec{n}_1}\En{\vec{n}_2}}_{h'h}
        =\frac{1}{8\pi}\sum_{k=0}^{3}
        \delta_{2\pi}\!\left(\alpha-\frac{k\pi}{2}\right)
        \begin{pmatrix}
            1&H_4 e^{4i\theta_1}\\
            H_4 e^{-4i\theta_1}&1
        \end{pmatrix},
    \end{align}
    and hence
    \begin{align}
        (c_2,c_3,c_4)=(0,0,1),\qquad |H_4|\le1.
    \end{align}
    In the centered-angle convention of \eqref{Eq:3DTwoPointSummary},
    $h(\alpha)=H_4\cos(2\alpha)f(\alpha)$.

    \item \textbf{Y events $\mathsf Y_a$.}
    These are three-atom events with reference directions $0,\vartheta,-\vartheta$
    and energy fractions $u,v,v$. The shape angle $\vartheta$ is distinct from
    the detector-pair angle $\theta=(\theta_1+\theta_2)/2$.
    \nopagebreak[4]
    \input{figs/events/y_event.tex}
    Define $a\equiv1+2\cos\vartheta$, with $0\le a\le1$.
    Energy and momentum conservation fix
    \begin{align}
        u=\frac{1-a}{3-a},\qquad v=\frac{1}{3-a}.
    \end{align}
    
    For the isotropic part we get 
    \begin{align}
        \frac{f(\alpha)}{2\pi}
        ={}&(u^2+2v^2)\delta_{2\pi}(\alpha)\nonumber\\
        &+2uv\bigl[\delta_{2\pi}(\alpha-\vartheta)
                   +\delta_{2\pi}(\alpha+\vartheta)\bigr]\nonumber\\
        &+v^2\bigl[\delta_{2\pi}(\alpha-2\vartheta)
                  +\delta_{2\pi}(\alpha+2\vartheta)\bigr].
    \end{align}
    Similarly, \eqref{Eq:extractingh} gives the off-diagonal contribution
    \begin{align}
        \frac{h(\alpha)}{2\pi}
        =q\Bigl\{&(u^2+2v^2\cos4\vartheta)\delta_{2\pi}(\alpha)\nonumber\\
        &+2uv\cos2\vartheta\bigl[\delta_{2\pi}(\alpha-\vartheta)
                                +\delta_{2\pi}(\alpha+\vartheta)\bigr]\nonumber\\
        &+v^2\bigl[\delta_{2\pi}(\alpha-2\vartheta)
                  +\delta_{2\pi}(\alpha+2\vartheta)\bigr]\Bigr\}.
    \end{align}
    Writing $r(a)=a+a^2-a^3=e_4$, we obtain
    \begin{align}
        (c_2,c_3,c_4)=\bigl(a^2,(1-a^2)^2,r(a)^2\bigr),
        \qquad H_4=q\,r(a).
    \end{align}
    \eqref{Eq:EventH4} gives $|H_4|\le r(a)$, with
    $q=H_4/r(a)$ for $a>0$. At $a=0$, the particles form the equilateral
    triangle and $h=H_4=0$. At $a=1$, the zero-energy particle drops out,
    leaving the back-to-back event.

    \item \textbf{X events $\mathsf X_r$.}
    Four atoms of energy $1/4$ form two back-to-back pairs whose axes
    lie at $\pm\vartheta/2$ relative to the vertical axis. The event is
    controlled by $r=\cos2\vartheta$, with $0\le\vartheta\le\pi/4$.
    \nopagebreak[4]
    \input{figs/events/x_event.tex}
    
    Defining
    \[
        D(\alpha)=\delta_{2\pi}(\alpha)+\delta_{2\pi}(\alpha-\pi),
    \]
    we get
    \begin{align}
        f(\alpha)&=\frac{\pi}{4}
        \bigl[2D(\alpha)+D(\alpha-\vartheta)+D(\alpha+\vartheta)\bigr],\\
        h(\alpha)&=\frac{\pi q}{4}
        \bigl[2rD(\alpha)+D(\alpha-\vartheta)+D(\alpha+\vartheta)\bigr],
        \qquad |q|\le1.
    \end{align}
    The harmonics are 
    \begin{align}
        (c_2,c_3,c_4)=\left(\frac{1+r}{2},0,r^2\right),
        \qquad H_4=q r,\qquad |H_4|\le r.
    \end{align}
    Here $e_4=r$, so $q=H_4/r$ for $r>0$. At $\vartheta=0$ ($r=1$),
    the pairs coincide and give the back-to-back event. At
    $\vartheta=\pi/4$ ($r=0$), $H_4=0$ but $q$ remains free and $h$ need
    not vanish; for example, $d_0=q/2$.
 
    \item \textbf{Two-square events $\mathsf D_r$.}
    Eight particles of energy $1/8$ form two squares rotated relative to
    one another by $\gamma$. Each square separately has zero momentum.
    \nopagebreak[4]
    \input{figs/events/two_squares.tex}
    The reference angles and shape parameter are
    \begin{align}
        \beta_{k,\pm}&=\frac{k\pi}{2}\pm\frac{\gamma}{2},
        \qquad k=0,1,2,3,\\
        r&\equiv\cos2\gamma,\qquad
        0\le\gamma\le\frac{\pi}{4},\qquad 0\le r\le1.
    \end{align}
    Define the ordinary and alternating square combs
    \begin{align}
        S(\alpha)&=\sum_{k=0}^{3}
          \delta_{2\pi}\!\left(\alpha-\frac{k\pi}{2}\right),&
        A(\alpha)&=\sum_{k=0}^{3}(-1)^k
          \delta_{2\pi}\!\left(\alpha-\frac{k\pi}{2}\right).
    \end{align}
    Summing the ordered pairs gives
    \begin{align}
        f(\alpha)&=\frac{\pi}{8}
          \bigl[2S(\alpha)+S(\alpha-\gamma)+S(\alpha+\gamma)\bigr],\\
        h(\alpha)&=\frac{\pi q}{8}
          \bigl[2rA(\alpha)+A(\alpha-\gamma)+A(\alpha+\gamma)\bigr],
          \qquad |q|\le1.
    \end{align}
    The harmonics are
    \begin{align}
        (c_2,c_3,c_4)=(0,0,r^2),\qquad
        H_4=q r,\qquad |H_4|\le r.
    \end{align}
    Here $e_4=r$, so $q=H_4/r$ for $r>0$. At $\gamma=\pi/4$ ($r=0$),
    the particles form a regular octagon: the terms in $h(\alpha)$ cancel, so
    $h(\alpha)=H_4=0$ for every $q$. At $\gamma=0$ ($r=1$), the two squares
    coincide and give the square event $\mathsf Q$, with $q=H_4$.
\end{itemize}

\subsection{The $(c_2,c_3)$-plane.}

We use the events before and their mixtures. For probabilities
$\omega_j\ge0$ with $\sum_j\omega_j=1$, their moments and source parameter
obey
\begin{align}
    c_n=\sum_j\omega_j|e_n^{(j)}|^2,
    \qquad |H_4|\le\sum_j\omega_j|e_4^{(j)}|.
    \label{Eq:3DEventMixture}
\end{align}
The second condition follows from the positive polarization weights
$|q|\le1$ in \eqref{Eq:EventH4}.

\begin{itemize}
    \item \textbf{$c_2\ge0$ and $c_3\ge0$.}
    These are diagonal entries of $\mathscr E^{(2,1)}\succeq0$ in
    \eqref{Eq:G}. The corresponding boundary
    segments are realized by
    \begin{align}
        c_2=0:&\quad
        \omega\mathsf T+(1-\omega)\mathsf Q,
        &&c_3=\omega,\quad 0\le\omega\le1-|H_4|,\\
        c_3=0:&\quad
        \omega\mathsf B+(1-\omega)\mathsf Q,
        &&c_2=\omega,\quad 0\le\omega\le1.
    \end{align}
    Only the square carries $H_4$ in the first set, both events
    can carry it in the second. The first segment therefore ends at
    $(0,1-|H_4|)$.

    \item \textbf{$c_2+c_3\le1$.}
    This follows from the rows $(i,m)=(0,0),(2,2),(3,3)$ of
    $\mathscr E^{(3,1)}$ in \eqref{Eq:G3}, giving the block
    \eqref{Eq:3Dc2c3}. It is saturated by
    \begin{align}
        \omega\mathsf B+(1-\omega)\mathsf T,
        \qquad (c_2,c_3)=(\omega,1-\omega),
        \qquad |H_4|\le\omega\le1.
    \end{align}
    Only the back-to-back event carries $H_4$, so the segment ends at
    $(|H_4|,1-|H_4|)$.

    \item \textbf{Curved bounds.}
    The relevant condition is $\mathscr E^{(1,0)}\succeq0$ in
    \eqref{Eq:E1}, giving, up to level 4
    \begin{align}
        |e_3|^2\le(1-|e_2|^2)^2,
        \qquad |e_3|^2\le(1-|e_4|)(1+|e_2|).
        \label{Eq:3DEventSchurBounds}
    \end{align}
    Eliminating the event modes and averaging with
    \eqref{Eq:3DEventMixture} gives, for $0<a<1$,
    \begin{align}
        c_3+(1-a)^2c_2
        &\le(1-a)^2+2a(1-|H_4|),
        \label{Eq:3D23RightSupports}\\
        c_3-\frac{(1-a)^3(1+a)^2}{a(a^2+a+2)}c_2
        &\le\frac{2(1+a)(1+a^2)}{a^2+a+2}(1-|H_4|).
        \label{Eq:3D23LeftSupports}
    \end{align}
    Here $a$ and $r(a)=a+a^2-a^3$ are the Y-event parameters of
    Appendix~A.1. For $0<|H_4|<1$, let $a_*$ satisfy $r(a_*)=|H_4|$.
    For $0<a\le a_*$, the right and left bounds are saturated,
    respectively, by
    \begin{align}
        &\omega\mathsf Y_a+(1-\omega)\mathsf B,
        \qquad \omega\mathsf Y_a+(1-\omega)\mathsf Q,
        \qquad \omega=\frac{1-|H_4|}{1-r(a)}.
    \end{align}
    Their coordinates follow by averaging the moments in Appendix~A.1.
    As $a\to0$, they join the straight segments above. At $a=a_*$,
    both mixtures become the single event $\mathsf Y_{a_*}$, giving
    the common kink
    \begin{align}
        (c_2,c_3,c_4)=\bigl(a_*^2,(1-a_*^2)^2,|H_4|^2\bigr).
        \label{Eq:3DYKinkSummary}
    \end{align}
    The same event inequalities fix the straight-segment cutoffs:
    $c_2=0$ implies $c_3\le1-|H_4|$, while $c_2+c_3=1$ allows
    only triangle and back-to-back moment values.
\end{itemize}

At $H_4=0$, these bounds give the triangle with vertices
$\mathsf Q=(0,0)$, $\mathsf B=(1,0)$ and $\mathsf T=(0,1)$.
At $|H_4|=1$, the region reduces to the segment joining $\mathsf Q$
and $\mathsf B$, with $c_3=0$. The limiting Y parameters are
$a_*=0$ and $a_*=1$, respectively.

\subsection{The $(c_2,c_4)$-plane.}

This region is fixed by the following inequalities and saturating events.
\begin{itemize}
    \item \textbf{$0\le c_2\le1$.}
    The lower bound is the mode-$2$ diagonal entry of
    $\mathscr E^{(2,1)}$ in \eqref{Eq:G}. The upper bound follows
    from the rows $(0,0),(2,-2)$ of $\mathscr E^{(2,0)}$ in
    \eqref{Eq:E2}, whose expectation value is the matrix
    $\bigl(\begin{smallmatrix}1&c_2\\c_2&1\end{smallmatrix}\bigr)$.
    The events $\mathsf D_r$, with $|H_4|\le r\le1$, saturate
    $c_2=0$; the back-to-back event $\mathsf B$ saturates $c_2=1$.

    \item \textbf{$c_4\le1$.}
    Use the rows $(0,0),(4,-4)$ of $\mathscr E^{(2,0)}$ in
    \eqref{Eq:E2}, giving the same two-by-two block with $c_2$
    replaced by $c_4$.
    It is saturated by
    \begin{align}
        \omega\mathsf B+(1-\omega)\mathsf Q,
        \qquad (c_2,c_4)=(\omega,1),\qquad 0\le\omega\le1.
    \end{align}

    \item \textbf{$c_4\ge|H_4|^2$.}
    This follows from the helicity block of
    $\mathscr E^{(2,1)}$ in \eqref{Eq:3DUnitarity}, with $J=n=2$.
    Mixtures of $\mathsf D_{|H_4|}$ and $\mathsf X_{|H_4|}$ saturate
    it and cover $0\le c_2\le(1+|H_4|)/2$.

    \item \textbf{$c_4\ge[\max(2c_2-1,0)]^2$.}
    The rows $(0,2,4)$ of $\mathscr E^{(1,0)}$ in \eqref{Eq:E1}
    give $|e_4-e_2^2|\le1-|e_2|^2$ on each configuration.
    Squaring the resulting lower bound on $|e_4|$ and applying
    Jensen's inequality after averaging gives
    \begin{align}
        c_4\ge\bigl[\max(2c_2-1,0)\bigr]^2.
        \label{Eq:3D24BoundsSummary}
    \end{align}
    The curved part is saturated by the X events,
    \begin{align}
        \mathsf X_r:\qquad (c_2,c_4)=\left(\frac{1+r}{2},r^2\right),
        \qquad |H_4|\le r\le1.
    \end{align}
    At $r=|H_4|$, this arc meets the horizontal boundary at
    $((1+|H_4|)/2,|H_4|^2)$. For $H_4=0$, mixtures of
    $\mathsf D_0$ and $\mathsf X_0$ also saturate the flat part,
    $c_4=0$ with $0\le c_2\le1/2$.
\end{itemize}
At $|H_4|=1$, only the segment $c_4=1$, $0\le c_2\le1$ remains.

\subsection{The $(c_3,c_4)$-plane.}

We use the same Y parameter $a_*$, fixed by $r(a_*)=|H_4|$.
\begin{itemize}
    \item \textbf{$c_3\ge0$.}
    This is the mode-$3$ diagonal entry of $\mathscr E^{(2,1)}$
    in \eqref{Eq:G}. The events $\mathsf D_r$ or $\mathsf X_r$,
    with $|H_4|\le r\le1$, saturate it and give
    $|H_4|^2\le c_4\le1$.

    \item \textbf{$c_4\ge|H_4|^2$.}
    This again follows from the $J=n=2$ helicity block of
    $\mathscr E^{(2,1)}$ in \eqref{Eq:3DUnitarity}.
    Mixtures of $\mathsf X_{|H_4|}$ and $\mathsf Y_{a_*}$ saturate
    this bound and cover $0\le c_3\le(1-a_*^2)^2$.

    \item \textbf{$c_3+c_4\le1$.}
    This follows from the rows $(i,m)=(0,0),(3,3),(4,4)$ of
    $\mathscr E^{(3,1)}$ in \eqref{Eq:G3}: they give the block
    \eqref{Eq:3Dc2c3} with $(c_2,c_3)$ replaced by $(c_3,c_4)$.
    It is saturated by
    \begin{align}
        \omega\mathsf T+(1-\omega)\mathsf B,
        \qquad (c_3,c_4)=(\omega,1-\omega),
        \qquad 0\le\omega\le1-|H_4|.
        \label{Eq:3D34BasicSummary}
    \end{align}
    One may replace $\mathsf B$ by $\mathsf Q$.
    Only this component carries $H_4$, so the segment stops at
    $(1-|H_4|,|H_4|)$. The event inequalities
    \eqref{Eq:3DEventSchurBounds} allow equality only at the triangle
    and $|e_4|=1$ moment values, fixing this cutoff.

    \item \textbf{$c_3\le(1-a_*^2)^2$.}
    The two Schur complements of $\mathscr E^{(1,0)}$ in
    \eqref{Eq:E1}, with rows $(0,1,2,3)$ and $(0,1,3,4)$, give
    \eqref{Eq:3DEventSchurBounds}. At fixed $|e_4|=r(a)$, they
    imply $|e_3|^2\le(1-a^2)^2$, saturated by $\mathsf Y_a$.
    This upper envelope is decreasing and concave in $|e_4|$,
    so averaging with \eqref{Eq:3DEventMixture} proves the stated
    bound. It is saturated by $\mathsf Y_{a_*}$, at the kink
    \begin{align}
        (c_3,c_4)=\bigl((1-a_*^2)^2,|H_4|^2\bigr).
    \end{align}

    \item \textbf{Remaining curved bounds.}
    The same rows of $\mathscr E^{(1,0)}$ in \eqref{Eq:E1},
    combined with \eqref{Eq:3DEventMixture}, give
    \begin{align}
        c_4+\lambda c_3+\mu|H_4|\le
        \max_{0\le a\le1}\left[
            r(a)^2+\lambda(1-a^2)^2+\mu r(a)
        \right],\qquad \lambda,\mu\ge0.
        \label{Eq:3D34SupportsSummary}
    \end{align}
    For each slope $\lambda$, minimize the resulting upper bound on
    $c_4+\lambda c_3$ over $\mu$. The tight bound is saturated by one or two Y events
    whose parameters maximize the bracket. For two such values
    $a,b$, the mixture $\omega\mathsf Y_a+(1-\omega)\mathsf Y_b$ has
    \begin{align}
        (c_3,c_4)
        &=\omega\bigl((1-a^2)^2,r(a)^2\bigr)
          +(1-\omega)\bigl((1-b^2)^2,r(b)^2\bigr).
    \end{align}
    Its weight obeys $\omega r(a)+(1-\omega)r(b)=|H_4|$ when
    the minimizing $\mu>0$. For $\mu=0$, the mean of $r$ may be
    larger than $|H_4|$, with the polarization weights $q$ setting
    the required source value. A single Y event gives the kink above.
\end{itemize}

At $H_4=0$, the region is the triangle with vertices
$\mathsf X_0=(0,0)$, $\mathsf T=(1,0)$ and $\mathsf B=(0,1)$.
At $|H_4|=1$, it reduces to $(0,1)$.

In particular, the origin of the $(c_3,c_4)$ plane is compatible
only with $H_4=0$, whereas the origin of the $(c_2,c_3)$ plane is
compatible with every $|H_4|\le1$.
These statements concern different projections and follow
directly from the event configurations above.

\section{Transverse spin-1 source in 4D}
\label{app:sec:illustration:spinone}

We specialize the constraints of Section~\ref{Sec:OuterMoments} to a
parity-invariant spin-one source. We use the normalization of
\eqref{eq:cl:incl}, so that
$c_\ell=\langle\sum_{ij}x_i x_jP_\ell(\vec n_i\cdot\vec n_j)\rangle$
and $c_0=1$. The brackets average over events; each event obeys
$\sum_i x_i=1$ and $\sum_i x_i\vec n_i=0$.

The Wigner--Eckart theorem allows only $\ell=0,1,2$ in the
one-point function, and momentum conservation removes $\ell=1$.
In the Cartesian source basis, \eqref{eq:onepointintro} at unit
energy has eigenvalues $(1-a_\cale)/(4\pi)$, twice, and
$(1+2a_\cale)/(4\pi)$. Local positivity, equivalently the full
matrix $\mathscr E^{(1,0)}$ in \eqref{eq:e104d}, therefore gives
\begin{align}
    -\frac12\leq a_\cale\leq1.
    \label{Eq:4DSourceRange}
\end{align}
The endpoints restrict the underlying events to be planar or collinear,
respectively, with their orientations averaged over.

For the two-point function in a fixed helicity state, parity leaves
$J=0,K=0$ and $J=2,K=0,\pm2$ in \eqref{eq:HJ0K}.
For example, the $J=2,K=0$ term is $H^2_{00}(z)P_2(\cos\Theta)$,
with
\begin{align}
    H^2_{00}(z)&=\sum_{m,\ell,\ell'}c^{(m)}_{\ell\ell'}
                         (A^2_{00})^{(m)}_{\ell\ell'},\nonumber\\
    (A^2_{00})^{(m)}_{\ell\ell'}
      &=(-1)^m\langle\ell m;\ell',-m|20\rangle
        \sum_\lambda\langle\ell\lambda;\ell',-\lambda|20\rangle
        Y^*_{\ell\lambda}(\theta/2,0)Y_{\ell'\lambda}(\theta/2,0).
    \label{Eq:4DSpinOneH20}
\end{align}
Angular momentum addition and parity allow only
$\ell'=\ell,\ell\pm2$; the $\ell=1$ rows vanish.
For illustration, retaining $\ell,\ell'\leq3$ gives
\begin{align}
    H^2_{00}(z)={}&\frac{\sqrt5}{8\pi}\operatorname{Re}c^{(0)}_{02}
                       [P_0+3P_1]\nonumber\\
    &+\frac5{112\pi}
      [2c^{(-2)}_{22}-c^{(-1)}_{22}-2c^{(0)}_{22}
                       -c^{(1)}_{22}+2c^{(2)}_{22}]
                       [P_0-3P_1-2P_2]\nonumber\\
    &+\frac1{96\pi}
      [-5c^{(-3)}_{33}+3c^{(-1)}_{33}+4c^{(0)}_{33}
                       +3c^{(1)}_{33}-5c^{(3)}_{33}]
                       [P_0-3P_1+5P_2+5P_3],
    \label{Eq:4DSpinOneTruncation}
\end{align}
where all $P_N$ have argument $\cos\theta=1-2z$.
Thus positivity of the fixed-$m$ blocks in \eqref{eq:chm:eachm}
constrains the spinning Legendre coefficients as well as the inclusive
ones. The following two projections admit explicit boundaries.

\subsection{\texorpdfstring{The $(c_2,c_3)$ plane}{The (c2,c3) plane}}
\label{App:4D23}
\label{App:4D23a09}

We give the boundaries in Fig.~\ref{fig:4DRegionsAnalytic} at
$a\equiv a_\cale$. For an event, let
$M=\sum_i x_i\vec n_i\vec n_i^{\,T}$.
A rank-one source weight with axis $\vec v$ gives
$\sum_i x_iP_2(\vec n_i\cdot\vec v)
=(3\vec v^{\,T}M\vec v-1)/2$.
Rotating this weight together with the event leaves the inclusive
multipoles unchanged. Decomposing positive source weights into
rank-one weights gives the event condition
\begin{align}
    \frac{3\lambda_{\min}(M)-1}{2}
       \leq a(\mathrm{Ev})\leq
    \frac{3\lambda_{\max}(M)-1}{2},\qquad
    \langle a(\mathrm{Ev})\rangle=a\,,
    \label{Eq:4DSourceInterval}
\end{align}
with $\lambda_{\text{min/max}}$ the minimum and maximum eigenvalues of $M$ for any event ``$\text{Ev}$''. The source condition is imposed on the ensemble $\braket{a(\text{Ev})}$, rather than separately on each event.

Only two nontrivial event families are needed. A deformed tetrahedron
has three equally spaced prongs at height $h=(1+2u)/3$, each of
energy $1/[3(1+h)]$, and an opposite axial prong of energy $h/(1+h)$.
An isosceles triangle has two prongs at height $h=2(1-v)/3$, each
of energy $1/[2(1+h)]$, and an opposite prong of energy $h/(1+h)$.
Here ``height'' denotes the dimensionless coordinate
$h=\vec n_i\cdot\hat z=\cos\theta_i$ of a prong's unit direction
on the sphere, where $\hat z$ is the chosen symmetry axis and
$\theta_i$ its polar angle. The opposite axial prong points along
$-\hat z$.

Their coordinates in the $(c_2,c_3)$ plane are
\begin{align}
    U(u)&=\left(u^2,\frac59(1-u^2-2u^3+2u^4)\right),
               &&-\frac12\leq u\leq1,\nonumber\\
    I(v)&=\left(\frac{1-2v+4v^2}{3},
                  \frac{40}{81}(1-v)^2(1+2v)^2\right),
               &&\frac14\leq v\leq1.
    \label{Eq:4D23AtomicCurves}
\end{align}
The source interval ends at $u$ for $U(u)$ (its upper end for $u\geq0$,
lower end for $u\leq0$), and at $v$ for $I(v)$.
The special cases are the tetrahedron $U(0)=(0,5/9)$,
the equilateral triangle $\mathsf T=I(1/4)=U(-1/2)=(1/4,5/8)$,
and the dijet $\mathsf B=U(1)=I(1)=(1,0)$.

The upper bounds below follow from $\mathscr E^{(1,0)}$ in
\eqref{eq:e104d}, evaluated on an event with harmonic rows
$\ell=0,1,2$. Taking the Schur complement of the constant and linear
polynomials, contracting the quadratic block and using the tracelessness
of the third moment gives
\begin{align}
    \left\|\sum_i x_i\vec n_i^{\otimes3}\right\|^2
       &\leq \operatorname{tr}M^2+(\operatorname{tr}M^2)^2
                         -2\operatorname{tr}M^3,\nonumber\\
    c_2(\mathrm{Ev})&=\frac{3\operatorname{tr}M^2-1}{2},\qquad
    c_3(\mathrm{Ev})=\frac52
                    \left\|\sum_i x_i\vec n_i^{\otimes3}\right\|^2.
    \label{Eq:4D23MomentBound}
\end{align}
At a fixed extremal eigenvalue of $M$, the resulting upper support
is convex in the product of the other two eigenvalues. Its maximum
therefore lies at an endpoint: two eigenvalues coincide, or one vanishes. These spectra are attained by the deformed tetrahedra and
isosceles triangles above. At positive source, an oblate endpoint lies
below a mixture of a prolate deformed tetrahedron and a triangle with
the same $c_2$ and source value.
Supporting inequalities for the remaining families thus hold for
every event, and averaging preserves them at fixed $a$.

\begin{itemize}
\item The straight bounds are
    \begin{align}
        a^2\leq c_2\leq1,\qquad c_3\geq0,\qquad
        c_3\leq\frac59+\frac5{18}c_2,\qquad
        c_3\leq\frac56(1-c_2).
        \label{Eq:4D23Lines}
    \end{align}
    Nonnegativity follows from $\mathscr E^{(2,1)}$ in
    \eqref{eq:4d:pov:21}. The source interval
    \eqref{Eq:4DSourceInterval} gives
    $a(\mathrm{Ev})^2\leq c_2(\mathrm{Ev})$; averaging and
    Cauchy--Schwarz give $a^2\leq c_2$. The upper bound on $c_2$ follows
    from $M\succeq0$ and $\operatorname{tr}M=1$; the last two
    inequalities follow from \eqref{Eq:4D23MomentBound}.
    Weighted orthogonal antipodal pairs attain $c_3=0$ for
    $a^2\leq c_2\leq1$. Mixing $U(a)$ with the antipodal event of
    the same $M$ gives the left edge $c_2=a^2$,
    $0\leq c_3\leq U_3(a)$.
    Tetrahedron--triangle mixtures attain the first upper line for
    $\max(a,-a/2)\leq c_2\leq1/4$, when this interval is nonempty.
    Triangle--dijet mixtures attain the second for
    $\max(a,1/4)\leq c_2\leq1$.

\item For $1/4<a<1$, the isosceles--dijet arc is
    \begin{align}
        c_3\leq\frac{5(1-c_2)^2(c_2+1-2a)}{6(1-a)^2},\qquad
        \frac{1-2a+4a^2}{3}\leq c_2\leq a.
        \label{Eq:4D23RightArc}
    \end{align}
    Its supporting lines follow from \eqref{Eq:4D23MomentBound}
    and \eqref{Eq:4DSourceInterval}. Equality is attained by
    \begin{align}
        \frac{1-a}{1-v}\,I(v)+\frac{a-v}{1-v}\,\mathsf B,
        \qquad \frac14\leq v\leq a,
        \label{Eq:4D23RightMixture}
    \end{align}
    with both events at their maximal source values.

\item The remaining upper arc for $0<a<1$ is attained by a mixture
    of a deformed tetrahedron and an isosceles triangle.
    Write $U=(U_2,U_3)$ and
    $I=(I_2,I_3)$. The common supporting plane is fixed by
    \begin{align}
        \det\begin{pmatrix}
            U_2'(u)&U_3'(u)&1\\
            I_2'(v)&I_3'(v)&1\\
            U_2(u)-I_2(v)&U_3(u)-I_3(v)&u-v
        \end{pmatrix}=0,
        \qquad \max(u,1/4)<v<\frac{1+u}{2}.
        \label{Eq:4D23Tangency}
    \end{align}
    This interval selects the unique supporting root $v=v(u)$;
    its endpoint values are $v(0)=1/4$ and $v(1)=1$.
    More explicitly, its inequality is
    \begin{align}
        c_3\leq\alpha+\kappa c_2-\gamma a,\qquad
        \gamma=\kappa U_2'(u)-U_3'(u),\qquad
        \alpha=U_3(u)-\kappa U_2(u)+\gamma u,
        \label{Eq:4D23LeftSupport}
    \end{align}
    where
    \begin{align}
        \kappa=
        \frac{I_3(v)-U_3(u)-U_3'(u)(v-u)}
             {I_2(v)-U_2(u)-U_2'(u)(v-u)},\qquad v=v(u).
    \end{align}
    On the branch in \eqref{Eq:4D23Tangency}, $\gamma\geq0$
    and the plane lies above both event families; the endpoint
    reduction in \eqref{Eq:4D23MomentBound} gives the exclusion.
    Equality at fixed $a$ is attained by
    \begin{align}
        (c_2,c_3)=\frac{v(u)-a}{v(u)-u}\,U(u)
                 +\frac{a-u}{v(u)-u}\,I(v(u)),
        \qquad 0\leq u\leq a\leq v(u).
        \label{Eq:4D23LeftArc}
    \end{align}
    The weights are nonnegative and their mean source value is $a$.

\item For $-1/2<a<0$, the remaining upper inequalities are
    \begin{align}
        \frac95c_3\leq
        1+\frac32t^2+2t^3
        +\left(\frac12+6t+6t^2\right)c_2
        +(3t+6t^2+4t^3)a,
        \qquad 0\leq t\leq\frac12.
        \label{Eq:4D23NegativeSupport}
    \end{align}
    They follow from the same moment bound, now taking the minimum
    source value in \eqref{Eq:4DSourceInterval}.
    On an oblate deformed tetrahedron $U(-s)$ the gap is
    $(t-s)^2(1/2-s)(3+4t+2s)\geq0$.
    On a prolate deformed tetrahedron $U(s)$ it is
    $2s^3(1-s)+(3/2+6t+6t^2)s^2
      -(3t/2+3t^2+2t^3)s+3t^2/2+2t^3\geq0$;
    the quadratic has nonpositive discriminant for $0\leq t\leq1/2$.
    Equality occurs at $U(-t)$ and $\mathsf T$. Hence the curved edge is
    \begin{align}
        (c_2,c_3)=\frac{a+1/2}{1/2-t}\,U(-t)
                   +\frac{-a-t}{1/2-t}\,\mathsf T,
        \qquad 0\leq t\leq-a.
        \label{Eq:4D23NegativeArc}
    \end{align}
    It joins $U(a)$ to the first upper line at $c_2=-a/2$.
\end{itemize}

At $a=0$ the region is the inclusive quadrilateral with vertices
$(0,0)$, $\mathsf B$, $\mathsf T$ and $U(0)$.
At $a=-1/2$ it is the triangle with vertices $(1/4,0)$,
$\mathsf B$ and $\mathsf T$; at $a=1$ it is the single point
$\mathsf B$. Mixing upper and lower boundary events at fixed
$(a,c_2)$ fills each vertical interval.
For the slice $a=0.9$ in Fig.~\ref{fig:fig_a09}, the deformed-tetrahedron corner
is $(81/100,221/9000)$, the isosceles kink is
$(61/75,392/10125)$, and the upper arc joins the straight line at
$(9/10,1/12)$.

\subsection{The inclusive and fixed-source bispectrum bands}
\label{App:4DBispectrum}

We use the bispectrum normalization of \eqref{eq:bispectrumdef}
and write
\begin{align}
    x=\frac{c_{222}}{c_2},\qquad y=\frac{c_{224}}{c_2},\qquad c_2>0.
    \label{Eq:4DBispectrumCoordinates}
\end{align}
In our normalization, the inclusive bounds of Ref.~\cite{Belin:2026wkc}
(see their Fig.~2) give the light-blue band in Fig.~\ref{fig:bispec_fig}:
\begin{align}
    -\frac12\leq x\leq1,\qquad
    \frac{35x^2-10x-7}{18}\leq y\leq\frac{5x+7}{12}.
    \label{Eq:4DBispectrumInclusive}
\end{align}
For the medium-blue band, fix $a_\cale=-\beta$ with
$0<\beta\leq1/2$. The horizontal range and upper line are unchanged,
while the lower bound becomes
\begin{align}
    y\geq
    \begin{cases}
        \displaystyle\frac{35x^2-10x-7}{18},
            &-\frac12\leq x\leq-\beta
                \quad\hbox{or}\quad 2\beta\leq x\leq1,\\[5pt]
        \displaystyle\frac{70\beta^3}{9(3\beta-x)}
                         -\frac59x-\frac7{18},
            &-\beta\leq x\leq2\beta.
    \end{cases}
    \label{Eq:4DBispectrumLower}
\end{align}
At the value $\beta=0.418$ used in the figure, the joins are
$x=-0.418$ and $x=0.836$. The left join is a corner and the right
join is tangent to the inclusive parabola. The inclusive band is
recovered as $\beta\to0$; at $\beta=1/2$ the hyperbola spans the
whole lower edge.

The inequalities and their saturating events can be obtained as follows.
For each event define
\begin{align}
    Q=M-\frac13\mathbb1,\qquad
    s=\operatorname{tr}Q^2,\qquad t=\operatorname{tr}Q^3,\qquad
    R=\sum_i x_i(\vec n_i^{\,T}Q\vec n_i)^2.
    \label{Eq:4DBispectrumMoments}
\end{align}
Angular integration in \eqref{eq:bispectrumdef} gives
\begin{align}
    c_2=\frac32\langle s\rangle,\qquad
    c_{222}=\frac92\langle t\rangle,\qquad
    c_{224}=\frac{35}{8}
        \left(\langle R\rangle-\frac47\langle t\rangle
                                  -\frac2{15}\langle s\rangle\right).
    \label{Eq:4DBispectrumMomentsToC}
\end{align}
In particular, the plotted coordinates are ratios of ensemble averages.

\begin{itemize}
\item The inclusive bounds follow from
    \begin{align}
        s^2\leq R\leq t+\frac{s}{3},\qquad
        6t^2\leq s^3,\qquad -\frac{s}{6}\leq t\leq\frac{s}{3}.
        \label{Eq:4DBispectrumEventBounds}
    \end{align}
    The first lower bound is the Schur complement of the
    constant and $\vec n^{\,T}Q\vec n$ rows of
    $\mathscr E^{(1,0)}$ in \eqref{eq:e104d}.
    The upper bound is the sum of positive integrals
    \begin{align*}
        \sum_{i<j}\big\langle[n_i(Qn)_j-n_j(Qn)_i]^2\big\rangle_{\rm Ev}
          =t+\frac{s}{3}-R\geq0,
    \end{align*}
    from the same matrix, with harmonic rows through $\ell=2$.
    The remaining bounds follow from the three
    eigenvalues of $Q$, with $M\succeq0$ and $\operatorname{tr}M=1$.
    Averaging gives
    \begin{align}
        \frac{\langle R\rangle}{\langle s\rangle}
          \geq6\left(\frac{\langle t\rangle}{\langle s\rangle}\right)^2,
        \qquad
        \frac{\langle R\rangle}{\langle s\rangle}
          \leq\frac{\langle t\rangle}{\langle s\rangle}+\frac13,
    \end{align}
    where the first inequality also uses Cauchy--Schwarz.
    Substitution in \eqref{Eq:4DBispectrumMomentsToC} gives
    \eqref{Eq:4DBispectrumInclusive}.

\item The central bound in \eqref{Eq:4DBispectrumLower}
    also uses the source interval in \eqref{Eq:4DSourceInterval}.

    Put $b=-3\lambda_{\min}(Q)/2$. Then
    \begin{align}
        0\leq b\leq\frac12,\qquad \langle b\rangle\geq\beta,
        \qquad t=bs-\frac89b^3.
        \label{Eq:4DBispectrumSource}
    \end{align}
    For any $x_0\in[-\beta,2\beta]$, set
    \begin{align}
        s_0=\frac{8\beta^3}{3(3\beta-x_0)},\qquad
        \mu=\frac{9s_0^2}{8\beta^3},\qquad
        L=\mu\left(\frac83\beta^2-s_0\right)\geq0.
    \end{align}
    The required supporting inequality follows from the event identity
    \begin{align}
        &s^2-\mu t+(\mu\beta-2s_0)s-L(b-\beta)\nonumber\\
        &\qquad=\left[s-s_0-\frac\mu2(b-\beta)\right]^2
          +(b-\beta)^2\left[\frac{8\mu}{9}(b+2\beta)
                                      -\frac{\mu^2}{4}\right]\geq0.
        \label{Eq:4DBispectrumCertificate}
    \end{align}
    For $s_0\leq64\beta^2/27$ the square bracket is nonnegative.
    For larger $s_0$ it is nonnegative if $b\geq\beta/4$, since
    $\mu\leq8\beta$. If $b\leq\beta/4$, the left-hand side instead
    reads $s^2+[\mu(\beta-b)-2s_0]s+8\mu b^3/9+L(\beta-b)$;
    every term is nonnegative, using
    $\mu(\beta-b)-2s_0\geq
       s_0[27s_0/(32\beta^2)-2]\geq0$.
    Replacing $s^2$ by $R$, averaging and using
    $\langle b\rangle\geq\beta$ gives
    \begin{align}
        \langle R\rangle-\mu\langle t\rangle
                    +(\mu\beta-2s_0)\langle s\rangle\geq0.
    \end{align}
    This is the tangent to the hyperbola at $x_0$; taking all
    $x_0$ proves the bound for arbitrary event mixtures.

\item The lower edges are attained by four equal-energy particles
    in the directions
    \begin{align}
        \vec n_{\eta_1,\eta_2}
          =(\eta_1\sqrt{\lambda_1},\eta_2\sqrt{\lambda_2},
                          \eta_1\eta_2\sqrt{\lambda_3}),
        \qquad \eta_1,\eta_2=\pm1.
        \label{Eq:4DBispectrumDisphenoid}
    \end{align}
    They have zero momentum, $M=\operatorname{diag}(\lambda_1,
    \lambda_2,\lambda_3)$ and $R=s^2$.
    For the inclusive parabola choose
    \begin{align}
        (\lambda_1,\lambda_2,\lambda_3)
          =\left(\frac{1+2x}{3},\frac{1-x}{3},\frac{1-x}{3}\right).
        \label{Eq:4DBispectrumParabolicEvents}
    \end{align}
    Its source interval contains $-\beta$ precisely on the two
    outer pieces of \eqref{Eq:4DBispectrumLower}.
    For the central arc choose $s_0$ at $x_0=x$ and
    \begin{align}
        \lambda_1=\frac{1-2\beta}{3},\qquad
        \lambda_{2,3}=\frac{1+\beta}{3}
                          \mathbin{\pm}\sqrt{\frac{s_0}{2}-\frac{\beta^2}{3}}.
        \label{Eq:4DBispectrumEigenvalues}
    \end{align}
    Taking the source axis along the first eigenvector gives
    $a_\cale=-\beta$.
    For either spectrum, the orthogonal antipodal event with energy
    $\lambda_k/2$ on each of $\pm\vec e_k$ has the same $M$ and
    $R=t+s/3$, and saturates the upper line.
    Mixing the two events at fixed $M$ fills the vertical segment.
    At $x=0$, the central construction with sufficiently small
    $\beta>0$ fills the inclusive vertical interval above its lower
    endpoint. That endpoint is the limit $c_2\to0$; the plotted
    inclusive band includes its closure.
\end{itemize}

These constructions establish both bands. The dark-blue region also
fixes the measured two-point multipoles and is not described by
\eqref{Eq:4DBispectrumLower}.

{\small
\bibliography{bibs}{}
\bibliographystyle{figs/utphys}}

\end{document}

%% file: figs/events/back_to_back.tex
\begin{center}
\begin{tikzpicture}[font=\small, line cap=round]
    \draw[gray!45] (0,0) circle (1.30);
    \draw[->,dashed,thick,orange!80!black] (0,-1.55)--(0,1.75);
    \node[anchor=west,align=left,text=orange!80!black]
        at (.14,1.53) {polarization\\axis};
    \foreach \b in {140,320} {
        \draw[->,thick,blue!65!black] (0,0)--(\b:1.30);
        \fill[blue!65!black] (\b:1.30) circle (2pt);
    }
    \fill (0,0) circle (1.2pt);
    \draw[->,gray!80] (140:.43) arc (140:320:.43);
    \node[fill=white,inner sep=1pt] at (230:.62) {$\pi$};
    \draw[->,gray!80] (90:.72) arc (90:140:.72);
    \node[fill=white,inner sep=1pt] at (115:.95) {$\psi$};
    \node[anchor=south east,align=right] at (140:1.42)
        {angle $\psi$\\$\beta=0,\quad x=\tfrac12$};
    \node[anchor=north west,align=left] at (320:1.42)
        {angle $\psi+\pi$\\$\beta=\pi,\quad x=\tfrac12$};
\end{tikzpicture}
\end{center}

%% file: figs/events/triangle.tex
\begin{center}
\begin{tikzpicture}[font=\small, line cap=round]
    \draw[gray!45] (0,0) circle (1.05);
    \draw[->,dashed,gray!75] (0,-1.30)--(0,1.45);
    \node[anchor=south] at (0,1.50) {$\hat z$};
    \foreach \b in {90,210,330} {
        \draw[->,thick,blue!65!black] (0,0)--(\b:1.05);
        \fill[blue!65!black] (\b:1.05) circle (2pt);
    }
    \fill (0,0) circle (1.2pt);
    \draw[->,gray!80] (90:.45) arc (90:210:.45);
    \node[fill=white,inner sep=1pt] at (150:.70) {$\frac{2\pi}{3}$};
    \node[anchor=south east,align=center] at (-.12,1.17) {$\beta=0$\\$x=\tfrac13$};
    \node[anchor=north east,align=center] at (210:1.17) {$\beta=\tfrac{2\pi}{3}$\\$x=\tfrac13$};
    \node[anchor=north west,align=center] at (330:1.17) {$\beta=\tfrac{4\pi}{3}$\\$x=\tfrac13$};
\end{tikzpicture}
\end{center}

%% file: figs/events/square.tex
\begin{center}
\begin{tikzpicture}[font=\small, line cap=round]
    \draw[gray!45] (0,0) circle (1.05);
    \draw[->,dashed,gray!75] (0,-1.30)--(0,1.45);
    \node[anchor=south] at (0,1.50) {$\hat z$};
    \foreach \b in {0,90,180,270} {
        \draw[->,thick,blue!65!black] (0,0)--(\b:1.05);
        \fill[blue!65!black] (\b:1.05) circle (2pt);
    }
    \fill (0,0) circle (1.2pt);
    \draw[->,gray!80] (90:.45) arc (90:180:.45);
    \node[fill=white,inner sep=1pt] at (135:.70) {$\frac{\pi}{2}$};
    \node[anchor=south east,align=center] at (-.12,1.17) {$\beta=0$\\$x=\tfrac14$};
    \node[anchor=east,align=center] at (-1.17,0) {$\beta=\tfrac\pi2$\\$x=\tfrac14$};
    \node[anchor=north east,align=center] at (-.12,-1.17) {$\beta=\pi$\\$x=\tfrac14$};
    \node[anchor=west,align=center] at (1.17,0) {$\beta=\tfrac{3\pi}{2}$\\$x=\tfrac14$};
\end{tikzpicture}
\end{center}

%% file: figs/events/y_event.tex
\begin{center}
\begin{tikzpicture}[font=\small, line cap=round]
    \draw[gray!45] (0,0) circle (1.10);
    \draw[->,dashed,gray!75] (0,-1.35)--(0,1.50);
    \node[anchor=south] at (0,1.55) {$\hat z$};
    \foreach \b in {90,198,342} {
        \draw[->,thick,blue!65!black] (0,0)--(\b:1.10);
        \fill[blue!65!black] (\b:1.10) circle (2pt);
    }
    \fill (0,0) circle (1.2pt);
    \draw[->,gray!80] (90:.45) arc (90:198:.45);
    \node[fill=white,inner sep=1pt] at (144:.68) {$\vartheta$};
    \node[anchor=south east,align=center] at (-.12,1.22) {$\beta=0$\\$x=u=\frac{1-a}{3-a}$};
    \node[anchor=north east,align=center] at (198:1.22) {$\beta=\vartheta$\\$x=v=\frac{1}{3-a}$};
    \node[anchor=north west,align=center] at (342:1.22) {$\beta=-\vartheta$\\$x=v=\frac{1}{3-a}$};
\end{tikzpicture}
\end{center}

%% file: figs/events/x_event.tex
\begin{center}
\begin{tikzpicture}[font=\small, line cap=round]
    \draw[gray!45] (0,0) circle (1.30);
    \draw[->,dashed,gray!75] (0,-1.45)--(0,1.60);
    \node[anchor=south] at (0,1.65) {$\hat z$};
    \foreach \b in {70,110,250,290} {
        \draw[->,thick,blue!65!black] (0,0)--(\b:1.30);
        \fill[blue!65!black] (\b:1.30) circle (2pt);
    }
    \fill (0,0) circle (1.2pt);
    \draw[<->,gray!80] (70:.70) arc (70:110:.70);
    \node[fill=white,inner sep=1pt] at (0,.92) {$\vartheta$};
    \node[anchor=south east,align=center] at (110:1.42) {$\beta=\tfrac\vartheta2$\\$x=\tfrac14$};
    \node[anchor=south west,align=center] at (70:1.42) {$\beta=-\tfrac\vartheta2$\\$x=\tfrac14$};
    \node[anchor=north east,align=center] at (250:1.42) {$\beta=\pi-\tfrac\vartheta2$\\$x=\tfrac14$};
    \node[anchor=north west,align=center] at (290:1.42) {$\beta=\pi+\tfrac\vartheta2$\\$x=\tfrac14$};
\end{tikzpicture}
\end{center}

%% file: figs/events/two_squares.tex
\begin{center}
\begin{tikzpicture}[font=\small, line cap=round]
    \draw[gray!45] (0,0) circle (1.35);
    \draw[->,dashed,gray!75] (0,-1.52)--(0,1.72);
    \node[anchor=south] at (0,1.77) {$\hat z$};
    \foreach \b in {108,198,288,18} {
        \draw[->,thick,blue!65!black] (0,0)--(\b:1.35);
        \fill[blue!65!black] (\b:1.35) circle (2pt);
    }
    \foreach \b in {72,162,252,342} {
        \draw[->,thick,orange!80!black] (0,0)--(\b:1.35);
        \fill[orange!80!black] (\b:1.35) circle (2pt);
    }
    \fill (0,0) circle (1.2pt);
    \draw[<->,gray!80] (72:.70) arc (72:108:.70);
    \node[fill=white,inner sep=1pt] at (0,.94) {$\gamma$};
    \node[anchor=south east,align=center] at (108:1.55) {$\beta=\tfrac\gamma2$\\$x=\tfrac18$};
    \node[anchor=south west,align=center] at (72:1.55) {$\beta=-\tfrac\gamma2$\\$x=\tfrac18$};
    \node[anchor=south east,align=center] at (162:1.48) {$\beta=\tfrac\pi2-\tfrac\gamma2$\\$x=\tfrac18$};
    \node[anchor=north east,align=center] at (198:1.48) {$\beta=\tfrac\pi2+\tfrac\gamma2$\\$x=\tfrac18$};
    \node[anchor=north east,align=center] at (252:1.55) {$\beta=\pi-\tfrac\gamma2$\\$x=\tfrac18$};
    \node[anchor=north west,align=center] at (288:1.55) {$\beta=\pi+\tfrac\gamma2$\\$x=\tfrac18$};
    \node[anchor=north west,align=center] at (342:1.48) {$\beta=\tfrac{3\pi}{2}-\tfrac\gamma2$\\$x=\tfrac18$};
    \node[anchor=south west,align=center] at (18:1.48) {$\beta=\tfrac{3\pi}{2}+\tfrac\gamma2$\\$x=\tfrac18$};
\end{tikzpicture}
\end{center}